\documentclass[twocolumn,preprintnumbers, showpacs, amsmath, amssymb]{revtex4}
\def\be{\begin{equation}}
\def\ee{\end{equation}}
\def\beq{\begin{eqnarray}}
\def\eeq{\end{eqnarray}}
\def\n{\nonumber}
\def\bay{\begin{array}}
\def\eay{\end{array}}

\begin{document}
\preprint{CIRI/03-smw04}
\title{Einsteinian Field Theory as a Program in Fundamental Physics}

\author{Sanjay M. Wagh}
\affiliation{Central India Research Institute, \\ Post Box 606,
Laxminagar, Nagpur 440 022, India\\
E-mail:cirinag\underline{\phantom{n}}ngp@sancharnet.in}

\date{April 18, 2004}
\begin{abstract}
I summarize here the logic that leads us to a program for the
Theory of the Total Field in Einstein's sense. The purpose is to
show that this theory is a logical culmination of the developments
of (fundamental) physical concepts and, hence, to initiate a
discussion of these issues.
\end{abstract}

 \maketitle
\newpage

\section{Introduction} \label{introduction}
Our purpose is to analyze in one place Einstein's reasoning that
led him to the program of a field-theory that, in his own words,
is: Continuous functions in the four-dimensional [continuum] as
basic concepts of the theory \cite{ein1}.

Einstein had summarized these reasons in \cite{ein1}. However, we
reconsider these from a different perspective, particularly, of
\cite{smw01}. Then, we also modify Einstein's program suitably.

Furthermore, von Laue summarized in \cite{laue} the developments
related to conservation postulates in Physics. In the present
article, we shall often use this excellent, historically as well
as scientifically important \footnote{Einstein commented
\cite{ein1} on this essay as: {\em Max von Laue: An historical
investigation of the development of the conservation postulates,
which, in my opinion, is of lasting value. I think it would be
worth while to make this essay easily accessible to students by
way of independent publication. ---}}, article to illustrate and
substantiate our physical arguments.

\section{Inertia, energy and conservation laws} \label{laue}

To begin with, and following von Laue, consider Galileo's famous
analysis leading to the concept of an inertia of a physical body.

Galileo observed that a body falling on the surface of the earth
from a certain altitude must obtain precisely that velocity which
it requires to return to its former level. Any deviation from this
law would but be able to furnish us a method for making the body
ascend by means of its own gravity, a conceivable {\em perpetuum
mobile}. We consider this to be impossible.

Now, following Galileo, let a body ascend, after it has fallen
downwards a certain distance, on an inclined plane. The lower the
inclination of the inclined plane toward the horizontal, the
longer the path the body requires to obtain its former level on
the plane. Galileo verified this experimentally. Then, if the
inclined plane were made horizontal, we may conclude that the body
will keep moving on it to infinity with undiminished velocity.

This last conclusion is {\em not\/} any experimental result but an
inference to be drawn by imagining an infinite horizontal plane
tangential to the surface of the earth, an obvious impossibility.

Now, continuing with the earlier analysis, the simplest form of
the law of inertia was obtained by Galileo, that the velocity of
each (force-free) body is maintained in direction and magnitude,
both. Here, the word {\em inertia\/} refers to the tendency of a
physical body to oppose a change in its state of motion. An {\em
inertial mass\/} is then a {\em measure\/} of the inertia of a
physical body.

Further, in those times, collisions were considered as the
simplest form of interaction of physical bodies. For $m$ as the
inertial mass and $v$ as the velocity of a body relative to an
observer, Descartes formed the quantity of motion, $m\,v$, the
linear momentum, and asserted that it is conserved in a collision.
He, however, considered velocity and, hence, also the linear
momentum as scalar quantities, for us today, an erroneous notion.

Huygens, on the other hand, realized correctly that the sum
(formed with correct signs) of $m\,v$ has the same value before
and after a perfectly elastic collision, and that the sum of
$m\,v^2$ is also conserved in such a collision.

Here entered Newton. He adopted the {\em geometric\/} approach in
defining appropriate physical quantities for physical bodies.
Newton therefore conceptualized velocity and, hence, linear
momentum also, as vector quantities. From this, Newton developed
(difference) {\em calculus\/} needed to deal with the notion of
velocity of a body as a tangent to its path.  On this basis,
Newton then proposed his famous three laws of motion.

The concept of a mass point was introduced by Newton as it was
needed in his geometric approach. (If not mass point, then what
would move with the tangential velocity along the one-dimensional
path?) We note, however, that, for Galileo, the notion of an
inertial mass was that of the measure of the inertia of a physical
body.

In Newton's {\em Principia}, two pronouncements appear: first, the
rate of change of vector of (linear) momentum of a {\em mass
point\/} per unit time equals vector of the force acting on it,
and the second, the forces between two mass points are equal and
opposite. Consequently, the interaction of an arbitrary number of
mass points never changes their total momentum, it remains
constant for any system not subject to external forces.

Then was formulated the law of conservation of angular momentum
which is one of the important consequences of the law of
conservation of linear momentum. In Newton's formulation, it is
also based on the concept of a mass point because the angular
momentum is a vector quantity: $\vec{\ell}=\vec{r}\times \vec{p}$
where $\vec{r}$ is the radius vector of the mass point relative to
a given fixed point, $\vec{p}=m \vec{v}$ is the vector of its
linear momentum and we use the cross product of vectors $\vec{r}$
and $\vec{p}$ to obtain the angular momentum vector $\vec{\ell}$
of the mass point.

Simultaneously, Newton also formulated his famous (inverse square)
law of gravitation, furnishing us the gravitational force of
attraction between two mass points. Here, Newton introduced a new
notion of the {\em source properties\/} of a particle. Precisely,
if $M$ is the {\em source or gravitational mass\/} of one particle
and $m$ is the inertial mass of another particle located at
distance $d$ from the first particle, then the gravitational force
of attraction (produced by $M$ and acting on $m$) is given by the
famous expression: \be
\vec{F}_g\,=\,-\,G\,\frac{m\,M}{d^2}\,\hat{d} \n \ee where $G$ is
Newton's constant of gravitation and $\hat{d}$ is a unit vector
along the line joining the two particles with origin of the
coordinates at the location of the gravitational mass $M$.

It is essential to {\em distinguish\/} between the {\em inertial
mass\/} and the {\em gravitational mass\/} of the newtonian
particle. These two are conceptions of very different physical
origins.

However, as first shown by Galileo's experiments at the leaning
tower of Pisa, the inertial and the gravitational mass of a
physical body are equal to a high degree of accuracy. That is an
experimental result. But, the fact that these two quantities {\em
are equal\/} is to be recognized as an {\em assumption\/} of the
newtonian theory. (This recognition played a crucial role in the
formulation of the General Theory of Relativity.)

Newton's three laws of motion and his law of gravitation then
provided us a complete solution to the problem of motions
(mechanics) of physical bodies as mechanical systems, as
collections of newtonian particles.

Here, the newtonian third law of motion, the law of equality of
action-reaction pair, and his law of gravitation show very
distinctly that his mechanics assumes action at a distance. Of
course, there is nothing objectionable in this and it is beside
the point as to whether Newton himself considered his law of
gravitation as some sort of approximation to be replaced
eventually by a law incorporating finite speeds of propagation.

On the basis of the mathematical formulation developed by him,
Newton could then calculate the planetary motions. Newtonian
mechanics obtained its major confirmation from Kepler's
astronomical observations. Moreover, all the other great
achievements of Newton's work, the theory of tides, the
equilibrium configurations of rotating bodies, the calculation of
the speed of sound etc.\ lent credence to various laws of
conservation, of linear momentum, of angular momentum and, hence,
also to this monumental newtonian formulation of mechanics of
physical bodies.

There however existed one problem with the newtonian framework,
that of optics - the theory of light. Newton's corpuscular theory
for light did not explain every phenomenon of light, and Newton
recognized this. In particular, the existence of umbra and
penumbra indicated that the light corpuscles penetrated
``forbidden'' region in the shadows behind objects illuminated by
light. Forbidden region exists as Newton's laws show that a
particle of light should not move there.

These problems of optical phenomena got neglected and did not
hinder in any way with the scientific developments of mathematical
character, of those times, in newtonian mechanics. Some, in
particular, Huygens, however considered these and developed the
wave theory of light which could explain the optical phenomena.

Primarily, one of the simplest forms of a general law of Nature is
to assert the conservation of some particular physical quantity.
That a given physical quantity is really subject to a conservation
principle is, however, to be decided only by experimentations.
Different experimentations then confirmed the conservation
principles obtainable from the newtonian mechanics.

That is why the impact of the newtonian formulation of mechanics
completely overshadowed the developments in Physics for the next
few centuries. That the mathematical structure of the newtonian
mechanics, erected by many others after Newton, required no new
experiments or observations is testimony enough to say that the
physical foundation laid by Newton was completely sufficient to
support it. This led the physicists of those times to (erroneous)
conviction that all of physics could be reduced to newtonian
mechanics.

Of course, the newtonian concept of a mass point had been the
basis of these developments because Newton's laws of motion
presuppose it. As we have already noted, this mass point is a
notion derived from the (point-wise) geometric approach to
(one-dimensional) paths of physical bodies. Then, we have also
noted earlier that the galilean notion of inertial mass is simply
that of the measure of the inertia of a physical body.

The concept of {\em kinetic energy}, $\frac{1}{2}m\,v^2$, was then
proposed and developed by Bernoulli (who gave us the word
``energy'') and Euler. That a change in this quantity in a closed
mechanical system did not at all result in a reduction in its
``capacity of action'' was emphasized by them.

The stimulus for a generalization of the concept of energy to
include other forms of ``energy'' existed from the experiences
with the thermal processes. That the kinetic energy or mechanical
work could be lost while the temperature of the bodies under
consideration increased was known. Thus, the fact that kinetic
energy could be transformed to heat was known and that led to
thermodynamical considerations of conservation postulates.

J R Mayer provided the reasonably accurate theoretical value for
the mechanical equivalent of heat and L A Colding obtained almost
the same value in his experiments involving friction.

Independently of Mayer, Helmholtz developed the principle of
conservation of energy and its implications. Helmholtz derived his
expressions for energy directly from the impossibility of the {\em
perpetuum mobile}. He, then, reached the concept of {\em potential
energy\/} for mechanical systems, of the potential energy of a
body experiencing gravitational force, of a charged body
experiencing electric force, of a magnetized body experiencing
magnetic force etc. His energy considerations applied to the
production of currents in galvanic cells, thermocouple,
electromagnetic induction etc.

Historically, Helmholtz's considerations were not generally
accepted in the beginning. However, G J Jacobi emphasized that in
them we essentially obtain the logical continuation of the earlier
ideas behind the science of newtonian mechanics. The initial
opposition to Helmholtz's ideas gradually disappeared and ``energy
considerations'' became important in Physics.

In the words of William Thomson (Lord Kelvin of Largs), we finally
had: We denote as energy of a material system in a certain state
the contribution of all effects (measured in mechanical units of
work) produced outside the system when it passes in an arbitrary
manner from its state to a reference state which has been defined
{\em ad hoc}. The words ``in an arbitrary manner'' embody the
physical law of the conservation of energy.

Thus, the law of conservation of energy was, gradually, found to
hold beyond the sphere of newtonian mechanics. The same could also
be called the situation with the law of conservation of linear
momentum, to begin with. It took a gradual while for this law to
emerge out of the sphere of newtonian mechanics but that needed
new conceptions in the form of {\em locality of actions}.

Although Newton's laws involved action at a distance, the
collisions of a specific mass point with the others in its
immediate vicinity in a mechanical system (of closely packed mass
points) are thinkable as {\em local\/} phenomena. A {\em local\/}
disturbance could then {\em propagate\/} in such a mechanical
system to other regions from the region of its origin. These
considerations lead us to the calculation of the speed of sound in
such {\em fluid\/} configurations. This is exactly how finite
speeds of propagation obtain in Newton's theory, although all its
basic laws are based on the action at a distance.

Concept of locality of action and related ideas developed from
those of {\em fluid properties\/} of matter. Considerations of
closely packed newtonian particles led to development of concepts
of density, pressure, fluxes, stresses etc.

We define density of newtonian particles (inertial masses) as a
function of space coordinates and integrate it over the volume
under consideration to obtain the inertial mass contained within
that volume. Motions of newtonian particles lead us to concepts of
momentum transfer, pressure, etc. Here, the mathematical procedure
for switching from the particle picture to the fluid picture is a
well-defined one, we note.

From the viewpoint of present mathematics, different physical
quantities are {\em measurable functions\/} defined over the
underlying [continuum] space. Measure Theory tells us as to how to
perform then the integration (averaging) procedures involved in
above mentioned considerations.

The principle of local action (as opposed to action at a distance)
and that of finite velocity of propagation of disturbances of
physical quantities, even in a vacuum, meaning a region with no
(newtonian) particles present in it, first gained prominence in
electromagnetism. (It is this connotation of the word vacuum that
is generally taken to be implied by it and will be used here.)

Considerations of static electricity prompted Coulomb to propose
the (inverse square) law of electrostatic force between two
charged bodies. Similar to the occurrence of
inertial/gravitational masses in Newton's law of gravitation,
Coulomb introduced the electric charge as a {\em measure/source of
the quantity of electricity\/} in his law of force between
electrically charged bodies. Similar considerations were also used
in magnetism.

The motion of a charged physical body along a (point-wise)
geometric path was again the pivotal concept behind these laws.
Therefore, of necessity, the electric charge became the intrinsic
property of a newtonian point-particle.

At this point, we therefore note that the newtonian particle is
then endowed with two distinct source attributes or properties,
{\em gravitational mass\/} and {\em electric charge}. The
gravitational mass acts as a source of its gravitational force
(defined as per Newton's law of gravitation) while the electric
charge acts as a source of its electric force (defined as per
Coulomb's law).

It then gradually emerged that the motion of a charged body
results into the (simultaneous) existence of its magnetic force
along with its electric force. Ampere's experiments and his laws
of magnetism associated with current (of charges) were the reasons
behind this realization.

J C Maxwell then connected all these empirical laws to provide a
sound mathematical foundation to the theory of electromagnetism.
Maxwell's theory then provided us the prediction of
electromagnetic radiation, an electromagnetic disturbance
propagating from the region of the source to other regions at the
speed of light.

Predictions of Maxwell's theory of electromagnetism were confirmed
in numerous experiments. In particular, contributions of Faraday
were noteworthy. Also, Hertz's spectacular confirmation of the
existence of electromagnetic radiation lent due credence to
Maxwell's theory.

For electromagnetic processes, Helmholtz's energy methods yield
merely a formula for the total energy. This is as long as one
believed in action at a distance without a transmitting medium.
But, the question of localization of action or disturbance then
lacked meaning.

It is against this background that Michael Faraday developed the
concept of {\em field\/} as the medium transmitting such localized
action or disturbance. The field was considered as a change in the
physical state of a system which was essentially located in the
dielectric. It is equally necessary to invoke the same conceptions
for even the empty space between the carriers of electric charge,
electric currents and magnets.

Note that Maxwell's theory does provide the energy density (of the
field) which is composed additively of an electric and a magnetic
term: $\frac{1}{8\,\pi}\,(E^2+B^2)$, where $E$ is the electric
field strength and $B$ is the magnetic field strength. It is a
necessary supplement of the field concept, field has energy
associated with, and inseparable from, it. The question therefore
arose of the nature of physical processes involving this energy of
the field of Faraday's conception.

J H Poynting then introduced the notion of a {\em flux\/} of
electromagnetic energy entirely on the basis of the mathematical
formalism of Maxwell's theory. He showed that there is a flux of
electromagnetic energy whenever an electric and a magnetic field
are present at the same time.

With this recognition, it is now possible to determine the route
by which the chemical energy, which in the galvanic cell is
transformed into electromagnetic energy, gets to wire that
completes the circuit, where that energy gets converted into heat.
Similarly, we can also trace the energy flow in a circuit
involving an electric motor that transforms it to mechanical work.

It was then recognized that the electromagnetic energy must also
exist in the space intervening its emitter and its absorber.
Emitter looses energy while the absorber gains energy only on the
absorbtion of radiation. Then, during the transit of the
electromagnetic radiation from the time of its emission to the
time of its absorbtion, the sum total of all energies can be
constant only if we take into account the energy of radiation.

Similar considerations also apply for the law of momentum. The
emitter of electromagnetic radiation experiences the opposite
force of the absorber of radiation. But, during the transit of
radiation, the electromagnetic radiation must carry the momentum
with it and ``deposit'' that momentum at the absorber. In fact,
Maxwell showed that a body which absorbs a light ray experiences a
force in the direction of that ray. This must also hold for all
electromagnetic fields.

Henri Poincar\'{e} showed that, if $S$ denotes the magnitude of
the flux of electromagnetic energy, then the field must contain
momentum of the magnitude $S/c^2$ per unit volume where $c$ is the
speed of light. Electromagnetic momentum was then shown to be
observable not only in phenomena with light and heat but also with
static fields.

As a matter of historical interest, this approach was the cause of
considerable difficulties and took a long time to gain acceptance.
The chief reasons behind the difficulties were the required
generalizations of the laws of conservation of energy and of
momentum.

Of certain importance is the conclusion of the inertia of
electromagnetic energy that follows from Poincar\'{e}'s expression
$S/c^2$. If we displace a carrier of electric charge, then the
motion of the corresponding electric field gives rise to a
magnetic field, and their coexistence leads both to a current of
energy and momentum. This additional momentum represents an
additional inertial mass in the system under consideration. For an
electron, this electromagnetic mass is of the same order of
magnitude as the observed mass.

The newtonian picture of a mechanical system of closely packed
mass points had always been at the background of these
electromagnetic considerations. Then, the question arises: what
constitutes field. It must be newtonian particles, each one of
some electromagnetic inertia, making up the medium or the field.
This medium was the {\em ether}. Interactions of these {\em
ethereal\/} particles would then provide us the mechanical
interpretation of Maxwell's equations.

Also, the ether was required to be incompressible since the
electromagnetic radiation was only of transverse type. But, if a
body moved through ether there must be observable effects of the
presence of ether on its motion. Such observational effects were
fruitlessly sought.

Ultimately, one got used to the concept of the ``field'' existing
independently of newtonian particles. Thus emerged the
field-particle dualism. A material particle in Newton's sense and
the (electromagnetic) field as a continuum existed side by side
with the material particle acting as a source of its field. The
newtonian (source) particle appeared here as a {\em singularity\/}
of the (electromagnetic) field it generated around it.

We owe this clear formulation of the field and the particle
dualism to H A Lorentz. In this formulation of Lorentz, the
newtonian action at a distance gets replaced by that of the field
which also represented radiation.

Disturbing here are two facts: firstly, kinetic energy (of a
newtonian particle) and the energy of the field appear as
physically unrelated entities, and secondly, the field energy
carried inertia but not that of a newtonian particle.

An obvious question is then of the nature of the inertia of the
field energy. But, it is thinkable that the inertia of field
energy is the same as the inertia of a newtonian particle. Then,
the concept of a newtonian particle would be simply that of a
region of special density of field energy. In that case one could
hope to deduce the concept of a particle and its equations of
motion from that of only the equations of the field.

Einstein then comments \cite{ein1} about these ideas as: {\em H A
Lorentz knew this very well. However, Maxwell's equations did not
permit the derivations of the equilibrium of the electricity which
constitutes a particle. Only other, non-linear field equations
could possibly accomplish such a thing. But no method existed by
which this kind of field equations could be discovered without
deteriorating into adventurous arbitrariness}.

In other words, Lorentz was aware of the above mentioned
disturbing facts and clearly recognized that the total inertia of
a newtonian particle could possess origin in the field conception.

However, the problem Lorentz faced was that of the linearity of
Maxwell's equations. Solutions of (linear) Maxwell's equations
obey superposition principle. Then, one could always superpose
required number of solutions to obtain the solution of any assumed
field configuration. The newtonian particle would still continue
to be the singularity of the final field configuration. Therefore,
there were no means here of removing all together the newtonian
particle that had the nature of the singularity of the field it
generated.

Some non-linear field equations could conceivably possess
singularity-free solutions for the field. Solutions of such
(non-linear) field equations would also not obey the superposition
principle. Then, one could hope that these (intrinsically
non-linear) equations for the (total) field would permit some
appropriate treatment of newtonian particles as singularity-free
regions of concentrated field energy. But, an obviously vexing
question was now that of the appropriate (non-linear) field
equations of this type. There did not exist (with Lorentz) any
physical guidelines (principles) for getting to these (non-linear)
field equations.

This however does not mean that such an approach is an
impossibility. In fact, it is a logical continuation of the
newtonian framework in a definite sense. This is what led Einstein
to express the following optimism.

Einstein remarks \cite{ein1} that: {\em In any case one could
believe that it would be possible by and by to find a new and
secure foundation for all of physics upon the path which had been
so successfully begun by Faraday and Maxwell. ---}

Then, we recall here that the galilean notion of (inertial) mass
of a physical body is that of the measure of the tendency of a
physical body to oppose a change in its state of motion. It is the
(point-wise) geometric approach of Newton that actually gave us
the concept of a mass point or a newtonian particle in Physics.

By giving up the point-wise geometric picture of Newton as basic
approach (but retaining it in some appropriate form), one may
still hope to treat inertia in the galilean sense of it being a
measure of the tendency of a physical body, an extended region of
the field energy, to oppose a change in its state of motion. This
is then the meaning of Einstein's above mentioned optimism.

The question is, of course, of reaching the (non-linear) field
equations of appropriate nature without venturing into meaningless
arbitrariness. We then need some definite physical principles to
reach to the non-linear field equations of the desired type. This
is what we turn to. However, we note that the route to appropriate
field equations had been long and difficult.

Now, as separate remarks, we firstly recall that the equality of
the inertial mass and the gravitational mass is certainly an
assumption of the newtonian mechanics. Newton's theory then does
not offer any explanation of this experimental result. This is
apart from the fact that Newton's theory does not provide
satisfactory explanations of the optical phenomena. Huygens's wave
theory of light then has existence separate, meaning independent,
from that of Newton's theory.

Secondly, from a geometric point of view, all coordinate systems
are among themselves logically equivalent. But, Newton himself had
realized (the bucket experiment) that the validity of his laws of
motion (for example, the law of inertia) is restricted to only
certain types of such reference systems, the {\em inertial\/}
frames of reference. Why this special status to inertial frames of
reference among all the possible others? This fact therefore
needed an explanation.

Newton's analysis (of the bucket experiment) then showed, quite to
his own dislike, that this explanation required the introduction
of {\em absolute space\/} as an omnipresent active participant in
all mechanical events, but the one which is not affected by the
masses and their motions. Else, his laws, in particular, the law
of inertia, could not have any physical content.

Newton himself could not resolve this impasse and nor could any
one else in his own times. Later on, it was also thought for some
while that the ether provided this absolute space of Newton's
theory. But, the ether, as a mechanical system made up of
newtonian mass points, must get affected by motions of masses.

It was then soon realized that, since the masses and their motions
did not affect the absolute space, there could not be any way of
establishing the existence of the absolute space. The absolute
space then came to possess the ghostly existence in the newtonian
framework.

Furthermore, in Newton's theory, there are essentially two
distinct concepts: firstly, that of the law of motion and,
secondly, that of the law of force. The law of motion is then
empty of content without the law of force. One may then be tempted
to ask if any (arbitrary) law of force could work. But, Newton's
law of gravitation and Coulomb's law of electrostatic attraction
possess, both, only a specific form: that of the inverse-square
type. The newtonian theoretical framework but provided no
explanation of this fact.

In essence, we therefore gradually came to recognize that various
of the basic concepts of the newtonian theoretical framework
needed to be replaced by suitable others.

\section{Special Relativity and Spacetime}
\label{specialrelativity}

It is then History that Einstein realized: many of the
simultaneous problems of electromagnetism and galilean principle
of relativity of newtonian mechanics could be resolved by invoking
a principle of the constancy of the speed of light in all the
(inertial) frames of reference.

Maxwell's equations do not permit a situation of spatially
oscillatory electromagnetic wave, the ``standing'' electromagnetic
wave. An electromagnetic wave always travels and, in vacuum, with
the speed of light. Moreover, various experiments did not indicate
the existence of any such standing electromagnetic wave. Then,
Maxwell's equations had experimental verification of certain
importance. Therefore, the galilean principle of relativity (the
galilean coordinate transformations) needed to be modified
suitably so that no standing electromagnetic wave could be
observed by any inertial observer.

That is to say, what are needed here are some ``new''
transformation laws (of coordinates) which keep Maxwell's
equations invariant. Then, no standing electromagnetic wave would
be obtained in all the (inertial) frames of reference if it cannot
be obtained in one frame of reference, that of Maxwell's standard
equations.

The coordinate transformations needed for this purpose were
already available to Einstein in the form of the Lorentz
transformations. This situation then led him to understand in a
clear manner what the spatial coordinates and the temporal
duration of events meant in the (new) formulation with the Lorentz
transformations.

Einstein's analysis led him to abandon the notion of the {\em
absolute simultaneity\/} of events and to the demand that the Laws
of Physics remain invariant under the Lorentz transformations.
This is similar in spirit to the demand of the newtonian theory
that the Laws of Physics be invariant under the galilean
transformations.

Minkowski's important contributions then made it possible to write
the Laws of Physics in a mathematical form that ensured their
invariance under the Lorentz transformations.

Now, to the criticism of the basis of the Special Theory of
Relativity.

Firstly, this theory provides us a (point-wise) geometric
description of physical bodies as mass points. This is evident
from the laws of motion in Special Relativity.

Secondly, the law of the force is also to be separately stated
and, hence, the relevant criticism of the newtonian mechanics also
applies.

Thirdly, as Einstein himself recognized \cite{ein1}, the special
theory of relativity treats, essentially independently, two kinds
of physical things, (1) measuring rods and clocks and (2) all
other things, like the electro-magnetic field, the material point
etc. This is unsatisfactory.

Surely, measuring rods and even clocks must be made up of material
points. Then, measuring apparatuses would also have to be
represented as objects consisting of moving material
configurations. Their treatment in Special Relativity is then not
a consistent one.

Next, the special status of the inertial frames of reference among
all the possible others remains unexplained in Special Relativity.
Therefore, the corresponding criticism of the newtonian theory
(absolute space) also applies to the theory of special relativity
as well.

Next, this theory does not explain the (observed) equality of the
inertial mass and the gravitational mass, although optical
phenomena are explained by way of Maxwell's equations in a
consistent manner in this theory.

Clearly, then, the theory of special relativity is not an entirely
satisfactory one since it leaves so much completely unexplained.

This, however, does not mean that it is not a step forward in the
direction of some satisfactory theory. There are certain
noteworthy and important achievements of Special Relativity.

In this theory, the {\em inertial mass\/} $m$ of a closed system
is identical with its energy $E$, {\em ie}, $E=m\,c^2$ where $c$
is the speed of light in vacuum. Further, the inertial mass $m$ is
dependent not only on the {\em rest mass\/} of the particle but
also on its velocity: \[m=m_o/ \sqrt{1- v^2/c^2}\] where $m_o$ is
the rest mass of the particle. This variation of $m$ with velocity
is a direct consequence of the Lorentz transformations.

Conceptually, if a physical body were to ``move'' in a ``field''
surrounding it, that body should ``display'' opposition to a
change in its state of motion and this opposition may be expected
to depend on its ``state of motion'' - the velocity that body
possesses. Recall that the inertial mass of a body is a measure of
its tendency to oppose a change in its state of motion.

As a consequence, the principles of conservation of linear
momentum and the conservation of energy are here fused into one
thing.

It also clearly recognizes that the coordinates have no absolute
physical meaning. This recognition is a very significant and
important step away from that of Newton's theory.

Anyway, it was very clear to Einstein \cite{ein1} that the Theory
of Special Relativity was only a transition from the newtonian
framework to incorporate electromagnetism in a consistent way.
That the special relativistic laws are linear laws obeying
superposition principle for their solutions was indication enough
for him that it is only a transitory phase in the formulation of
an appropriate (non-linear) theory of the field.

However, what emerged in the special theory of relativity is the
Minkowskian four-dimensional continuum indicating clearly that the
coordinates have no absolute physical meaning. But, differences of
coordinates had some definite physical meaning in this theory,
that of physical length and of duration of physical time.

It then gradually became clear to Einstein \cite{ein1} that even
the differences of coordinates need not possess any absolute
physical meaning. It was not any ``easy to come by'' realization.
This is what we turn to next.

\section{General Relativity and Curved Spacetime}
\label{generalrelativity}

There were many problems with the theory of special relativity and
one could easily have chosen to rectify them individually starting
from different possible perspectives. However, as is well known,
Einstein chose to concentrate only on one of them: the problem of
the equality of the inertial and the gravitational mass.

This fact is {\em not\/} just mere luck. Einstein's line of
argument has been clearly stated \cite{ein1} by him as the
following one. Firstly, one can see very clearly that, in the
special theory of relativity, if the inertial mass depended on the
velocity of a material point, then its gravitational mass would
also depend on its velocity (kinetic energy) in exactly the same
manner if, of course, the equality of the inertial and the
gravitational mass holds. This equality is but known
experimentally to be true to a high degree of accuracy.

Then, the {\em weight\/} of a physical body would depend on its
{\em total energy\/} in a precise manner since the inertial mass
now includes also the kinetic energy of the body. In essence, the
acceleration of a material system falling freely in a given
gravitational field is then independent of the nature (total
energy) of the falling system.

It then occurred to Einstein \cite{ein1} that: {\em In a
gravitational field (of small spatial extent) things behave as
they do in a space free of gravitation, if one introduces in it,
in place of an ``inertial system,'' a reference system which is
accelerated relative to an inertial system}.

The above is a statement of the {\em equivalence principle\/},
that represents the equality of the inertial and the gravitational
mass. Then, in a real gravitational field, it is therefore
possible to regard an appropriate reference frame, the frame of
reference of a falling body, as an ``inertial'' frame of
reference. The concept of the inertial frame becomes completely
vacuous, then.

The equivalence principle then implies that the demand of the
Lorentz invariance of physical laws is too narrow. Therefore, it
is  necessary to demand that the physical laws be invariant
relative to non-linear transformations of coordinates in the
four-dimensional continuum. This is then the principle of general
covariance.

Locally, meaning in a small region of the four-dimensional
continuum, the laws of special relativity must, however, hold in
an approximation. In the context of modern terminology, a general
four-dimensional spacetime manifold is then required to be locally
Lorentzian. The Lorentz group is therefore a subgroup of the group
of general coordinate transformations.

Einstein then raised \cite{ein1} the following two questions: {\em
Of which mathematical type are the variables (functions of the
coordinates) which permit the expression of the physical
properties of the space (``structure'')? Only after that: Which
equations are satisfied by those variables?}

Interestingly, he writes below these questions in \cite{ein1}
that: {\em The answer to these questions is today by no means
certain}. (That was in 1949, some decades after the (standard)
field equations of general relativity were given by him!)

We now trace here the path \cite{ein1} that Einstein chose to
reach the standard (Einstein) field equations of general
relativity.

Although we do not, to begin with, know which mathematical type
are the variables of the theory, we do know with certainty one
special case, the case of ``no gravitational field'' - the
spacetime of the special theory of relativity.

The line element of the Minkowski space is \[
ds^2=-\,dx_1^2+dx_2^2+dx_3^2+dx_4^2 \] for a properly chosen
coordinate system. It is a {\em measurable distance between
events}.

Now, referring to an arbitrary coordinate system, the same
(Minkowski) line element (metric) is expressible as
\[ds^2=g_{_{ij}}dx_idx_j\] where $g_{_{ij}}$ are the (real) metric
coefficients forming a symmetric tensor of rank two and $i,j$ take
values from 1 to 4. (Here, we have also used the (Einstein)
summation convention.)

If the first derivatives of $g_{_{ij}}$ do not vanish after the
coordinate transformation, then there exists a {\em special\/}
gravitational field in the (new) coordinate system in the sense
that it is ``accelerated'' with respect to the earlier coordinate
system.

Detailed mathematical investigations of metric spaces were carried
out by Riemann and various relevant conceptions existed before
Einstein began his search for the variables of the physical
properties of the space structure.

From Riemann's investigations of metric spaces, the Minkowski
spacetime can be uniquely characterized as the one for which the
Riemann curvature tensor vanishes. That is, the Minkowski
spacetime is a {\em flat\/} manifold. Then, it can also be seen
easily that the path of a mass point (not acted upon by any force)
in the Minkowski spacetime is a straight line and also a geodesic
of the Minkowski spacetime. A geodesic (straight line) is then
already a characterization of the Law of Motion in the (flat)
Minkowski spacetime.

In Einstein's words \cite{ein1}: ''The {\em universal\/} law of
physical space must now be a generalization of the law just
characterized.''

He then assumes that there are two steps of generalization,
namely, \begin{description} \item{(a)} pure gravitational field
\item{(b)} general field (in which quantities corresponding
somehow to the electromagnetic field occur, too).
\end{description}

The situation of (a), that of the pure gravitational field, then
is characterized by a symmetric (Riemannian) metric (tensor of
rank two) for which the Riemann curvature tensor does not vanish.
Now, if the universal field law is required to be of the second
order of differentiation and also linear in the second derivatives
of the metric coefficients $g_{_{ij}}$, then only the vanishing of
the Ricci tensor comes under consideration as an equation for the
(vacuum) field.

It may also be noted that the geodesics of the metric can then be
taken to represent the law of motion of the material point.
Therefore, the law of motion (of a mass point) is already
incorporated in that for the field.

Einstein \cite{ein1} then goes on elaborating further: {\em It
seemed hopeless to me at that time to venture the attempt of
representing the total field (b) and to ascertain field-laws for
it. I preferred, therefore, to set up a preliminary formal frame
for the representation of the entire physical reality; this was
necessary in order to be able to investigate, at least
preliminarily, the usefulness of the basic ideas of general
relativity.}

Now, let us therefore consider, following Einstein \cite{ein1},
this preliminary (formal) formulation of the (non-linear) field
theory based on the principle of general covariance.

Then, in Newton's theory, the gravitational field obeys the law
(Poisson's equation): \[ \nabla^2 \Phi = 4\pi\,G\,\rho \] where
$\Phi$ denotes the gravitational potential and $\rho$ denotes the
density of the sources generating that gravitational potential.

In general relativity, it is the Ricci tensor which takes the
place of $\nabla^2\Phi$. Therefore, Einstein proposed that the
preliminary formulation of general relativity be based on the
equations \[ R_{_{ij}}-\frac{1}{2}\,R\,g_{_{ij}}=-\,\kappa\,
T_{_{ij}}\] where $R_{_{ij}}$ denotes the Ricci tensor, $R$
denotes the Ricci scalar, $\kappa$ denotes a proportionality
constant and, importantly, $T_{_{ij}}$ denotes the energy-momentum
tensor of matter. In so far as these equations are concerned, the
energy-momentum tensor does not contain the energy (or inertia) of
the pure gravitational field.

The form of the left hand side of these equations (geometric part)
is chosen to be such that its divergence in the sense of absolute
differential calculus vanishes since a similar divergence of the
right hand side (matter part) must vanish from conservation
principles.

In this connection, Einstein expresses \cite{ein1} his judgement
and concerns about these preliminary equations as: {\em The right
side is a formal condensation of all things whose comprehension in
the sense of a field theory is still problematic. Not for a
moment, of course, did I doubt that this formulation was merely a
makeshift in order to give the general principle of relativity a
preliminary closed expression. For it was essentially not anything
{\em more\/} than a theory of the gravitational field, which was
somewhat artificially isolated from a total field of as yet
unknown structure}.

Then, keeping in mind these important remarks, let us consider
these two preliminary formulations of the above field equations.

It is then History that within a few months of Einstein's
publication of these field equations, K Schwarzschild obtained
(under heroic circumstances) the first solution of these highly
non-linear differential equations in a spherically symmetric case
of pure gravitational field. The Schwarzschild solution, a
spherically symmetric spacetime geometry, represents the exterior
or the vacuum gravitational field of a point mass.

Further solutions of these field equations (and history of some)
can be found in \cite{stdworks}.

However, noteworthy for us here are also the Kerr-Newman
spacetimes which represent a mass point with electric charge and
spin. Of course, when the electric charge and the spin are
vanishing, the Kerr-Newman spacetimes reduce to the Schwarzschild
spacetime.

We shall refer to this ``preliminary formulation'' as the
``standard general relativity'' since, over the decades, it has
indeed become one of the standard formulations of general
relativity.

Since our purpose here is not to consider the history of these
developments but rather that of considering, partially, the
history of the development of associated ideas, we shall now turn
to the criticism of this ``preliminary formulation'' or the
standard general relativity.

\section{Criticism of Standard General Relativity}
\label{criticismofgr}

We recall that Einstein himself was never at ease with this
standard general relativity. In \cite{ein1}, he says: {\em If
anything in the theory as sketched - apart from the demand of the
invariance of the equations under the group of the continuous
coordinate transformations - can possibly make the claim to final
significance, then it is the theory of the limiting case of the
pure gravitational field and its relation to the metric structure
of space.}

However, ``matter'' is not any part of the theory of the
``vacuum'' or pure gravitational field. Einstein, of course,
recognized this but hinted only that it is the relation of the
metric structure with the field which holds in this case can
possibly make the claim to final significance.

Since physical objects must be made up of material particles,
there simply cannot be physical objects, hence, measuring rods and
clocks, in this case of the pure gravitational field. Therefore,
the criticism that Einstein himself levelled against the special
relativity applies here.

It also does not therefore  make ``real sense'' to say that the
geodesics of the field provide the law of motion for the material
points since these are {\em not\/} the part of the equations of
the pure gravitational field. However, such a relation may be
expected in the ``final theory'' of the general field, the case
(b) referred to earlier.

Because of this problem, we also cannot claim that the equivalence
principle has been incorporated in the theory of the pure
gravitational field. That is, the equality of the inertial and the
gravitational mass is {\em not\/} explainable on the basis of the
theory of the pure gravitational field. This fact then invites the
same criticism as that of the special theory of relativity.

Furthermore, because of the same problem, it also cannot be
claimed that the formalism incorporates the law of force, which
must be separately stated. Then, again, this fact invites the same
criticism as that of the special theory of relativity and of the
newtonian theory.

Next, the inertia associated with the field energy is also not the
inertia of the material particle in this framework. This is then
very similar to the case of Maxwell's electromagnetism. Therefore,
the relevant criticism of the field-particle dualism in that
theory also applies to this case of the pure gravitational field.

But, what is more offending here is that, in this case of the pure
gravitational field, there definitely are the laws for the field,
but there simply cannot be laws for the material particles
generating that gravitational field.

To see this, notice that a mass point in this case is,
necessarily, a {\em curvature singularity\/} of the spacetime
manifold.

A point-particle has existence only at one spatial location and
everywhere else, except at this {\em special\/} location, it is
the non-singular gravitational field of the particle that
``exists''. In its spherically symmetric spacetime {\em geometry},
Schwarzschild or the Reissner-Nordstrom geometry, all the points,
except for the location of the mass-point, are then ``equivalent''
to each other in the sense of matter. That is, there does not
exist a material particle at any of these points.

This gravitational field only diverges at the location of the
mass-point unless, perhaps, we include the self-field in some way,
which, of course, is not the case with the equations of the pure
gravitational field. Hence, this {\em geometry\/} should not
``include'' the location of the mass-point.

Then, any mass-point is a curvature singularity of the manifold
describing its ``exterior'' gravitational field which, like in the
newtonian case, diverges at the mass point.

Addition of other source attributes of a physical body, like
charge, spin etc., does not change this situation. The Kerr-Newman
family is obtainable (in the case of the pure gravitational field)
and the point-mass with charge and spin is a spacetime singularity
in it.

The issue of the inadequacy of such a description of a physical
body would have arisen if the Kerr-Newman family of spacetimes
were not obtainable for the case (a) of the pure gravitational
field. We could then have said that the pure mass-point is an
inadequate description of a ``physical body'' and the addition of
other attributes ``takes us away from the case (a)'' to some new
situation, the case (b) of the general field, with some new
description of a physical body.

Restricting to the case (a) of the pure gravitational field, we
then note that, strictly speaking, a path of a mass point is not a
geodesic since the mathematical structure defining a geodesic
breaks down at every point along the path of the singularity. We
may try to circumvent this problem in the following ways.

We may surround the singular trajectory of a point-particle by an
appropriately ``small'' world-tube, remove the singular trajectory
and call this tube the ``geodesic'' of the particle. But, the new
spacetime is {\em not\/} the original spacetime and, by the
physical basis of general relativity, it corresponds to different
gravitational source. Therefore, this procedure is not at all
satisfactory.

We may, along a geodesic, change the spatial coordinate label(s),
by which we ``identify the singularity'', with respect to the time
label of the spacetime geometry and may term this change the
``motion'' of a particle along a geodesic. However, this is only
some ``special'' simultaneous relabelling of the coordinates. But,
a singularity in the spacetime does not ``move'' at all.

In the case (a) of the pure gravitational field, the motion of a
particle must be a ``singular trajectory'' in the spacetime if a
particle, the singularity, ``moves'' at all. It is also equally
clear that any singular trajectory is {\em not\/} a geodesic in
the spacetime since such a trajectory is {\em not\/} the part of
the smooth spacetime geometry.

Then, strictly speaking, geodesics of a spacetime do {\em not\/}
provide the {\em law of motion\/} of particles. Thus, strictly
speaking, we have to specify {\em separately\/} the law of motion
for the particles. It is also clear that the equations of pure
gravitational field do not (strictly speaking, cannot) specify
this law of motion for the particles.

But, mathematically speaking, there simply cannot be any laws for
the motion of a geometric singularity! That is why, in the case
(a) of the pure gravitational field, we have only the (non-linear)
equations for the field but no equations of motion for the sources
of that field.

In Newton's theory, the laws of motion for the particles were
Newton's laws of motion, but there were none for the field. The
field laws had to be postulated separately in the newtonian
framework and it was possible to formulate such laws for the field
independently. This is what Maxwell's electromagnetism achieved.

Then, the case (a) of the pure gravitational field leads us to a
situation which is the other extreme of that of Newton's theory.
We have the laws for the field in this case and there are none for
the particles. But, it is more offending that, with it, we cannot
even formulate separate laws for the motion of the mass points or
particles.

To further clarify this situation, let us consider a mass point
executing a simple harmonic motion along a straight line, the
$X$-axis, say, in relation to two other mass points located some
distance away from the first one along the $X$-axis. At $t=0$, let
the first mass point be located at, say, $x=0$ and the second one
at, say, $x=x_2$ and the third one at, say, $x=x_3$.

To be able to describe the simple harmonic motion of the first
mass point in relation to other mass points under consideration,
we need a spacetime having two singular curves at $x=x_2$ and at
$x=x_3$ which run parallel to the time axis (in the Minkowski-type
diagram) and having another singular curve, of sinusoidal
character, running along the time axis.

It is then thinkable that such a solution of the field equations
of the pure gravitational field exists and is obtainable
explicitly. This is however not the issue under consideration
here.

The issue is of the ``cause'' behind the harmonic motion of a
curvature singularity, and there cannot be any laws providing us
this cause of the motion. The solution to the field equations of
pure gravitational field is then ``ghostly.''

That is why, if we continue to consider, with complete disregard
to this fundamental difficulty, the case (a) of the pure
gravitational field, we are bound to face inconsistencies. That
this is indeed the case can be seen as under.

Further, if there existed other particles, then the spacetime
geometry would be {\em different\/} from that of a single
point-particle, now having singularities at every location  of
particles in it. Then, a physical body is, for our considerations,
a collection of curvature singularities.

Now, we may want to replace a collection of particles by that of
their smooth (fluid) distribution. To achieve this, let us attach
a ``weight function'' to each mass-point in our collection. Each
of these weight-functions must ``diverge at a suitable rate'' at
its mass-point to balance the field-singularity at that location.
Then, the resulting ``weighted-sum'' should provide a smooth
``volume-measure'' as well as a smooth source density.

Mathematically, $d\mu = \rho(x) \,d\mu_o$, where $\mu$ is the
volume-measure of the smooth geometry, $\mu_o$ that of the
geometry with singularities and $\rho(x)$ is the required source
density.

But, any smooth $\rho(x)$ is {\em impossible\/} if the geometry of
$\mu_o$-measure has curvature singularities. To keep $\mu$ smooth,
$\rho(x)$ must diverge at the singularities. Thus, associated
difficulties arise for constructing a smooth spacetime here.

Now, ignore also these difficulties and consider a spherical star,
using the fluid approximation, with a smooth spherical spacetime.

Consider two copies of such a star separated by a very large
distance today. Now, the spacetime of two stars taken together is
{\em not\/} globally spherically symmetric even though the
spacetimes of single stars are globally spherically symmetric.
However, the evolution of each star would be ``weakly'' affected
by the other distant star, an expectation justifiable on general
considerations.

Let stars collapse. Then, if we had some definite outcome in the
collapse of the star, exactly the same outcome is observable at
locations of each star in the above (symmetric) situation.

We may add further copies of the same star, separated by the {\em
same\/} large distance from original stars and from each other. If
stars collapse, the same end-result will be seen for each of the
stars in this, arbitrary, situation too.

Now, if gravitational collapse leads to a naked singularity for
the spherical spacetime, two totally {\em inequivalent\/}
descriptions obtain for a particle, that it is sometimes naked and
is sometimes covered by a horizon, an inconsistency.

There are then enough indications for us here that the case of the
pure gravitational field is an internally inconsistent one to
consider.

We then note that Einstein, in recognition of some of these
problems, wrote in (\cite{ein1}, p. 675) that: {\em Maxwell's
theory of the electric field remained a torso, because it was
unable to set up laws for the behavior of electric density,
without which there can, of course, be no such thing as an
electromagnetic field. Analogously the general theory of
relativity furnished then a field theory of gravitation, but no
theory of the field-creating masses. (These remarks presuppose it
as self-evident that a field-theory may not contain any
singularities, i.e., any positions or parts in space in which the
field-laws are not valid.)}

However, any considerations of the pure gravitational field in
general relativity provide us the equations for the field but, in
complete contrast to Maxwell's theory, there clearly can never be
any, not even faintest, possibility of our formulating any
dynamical laws for the sources of that field, may those be even
independent of these field laws. These considerations provide us,
similar to that of Newton's absolute space, not just a mere torso
but a ghost \footnote{Therefore, just as various theoretical
conclusions based on the existence of the absolute space of
Newton's theory were unjustified, destined to be failure from the
very beginning, the solutions to the equations of the pure
gravitational field are similarly not justifiable. Clearly, a
black hole is one of the many ``children'' of the ``ghost'' of the
field equations of the pure gravitational field and, hence, is
itself a ghost. Similar is the case with a ``naked singularity.''  \\
Then, anything done employing such ``children of the ghost'' is
dubious. For example, the works of the author in ``The energetics
of black holes in electromagnetic fields by the Penrose process''
(Wagh, S.M. and Dadhich, N.) {\it Physics Reports}, {\bf vol.
183}, p. 137 (1989) (and references therein) are, simply, {\em
dubious}, since these works use this ``ghost'' of the equations of
the pure gravitational field in every possible manner. A naked
singularity, it too being dubious, is then not useful for any astrophysical purposes.\\
It is then evident that efforts (for example, see, McClintock J E,
Narayan R, Rybicki G B (2004) {\bf Database: astro-ph/0403251} and
references therein) to establish the existence of a black hole
horizon on the basis of observations of astronomical sources are
not justifiable. The corresponding observations must then admit
explanations of some entirely different nature. } for us to deal
with.

Recall that, since the masses and their motions did not affect
Newton's absolute space, no means are possible of establishing the
existence of the absolute space. Then, the absolute space has the
ghostly existence in the newtonian framework. Similarly, the field
equations of the pure gravitational field also posses a ghostly
existence since the laws of motions of the sources of the pure
gravitational field are an impossibility here.

Without the laws for the motion of sources (generating the field
under consideration), there is no meaning to the equations of the
pure gravitational field because these equations for the field do
not permit any understanding of the physical situations, as the
above example shows.

Now, we may be tempted to imagine that the solutions of the
equations of the pure gravitational field will be able to
approximate the ``true'' situation in some useful way.

Unfortunately, this optimism ignores some very important aspects
of the general theory of relativity, those related to the fact
that its solutions do not follow any superposition principle. It
may then be noticed that the geometry with (smooth) matter fields
cannot be approximated by the collection of those representing the
pure gravitational field since the latter would possess the
curvature singularities at the locations of sources in it while
the former geometry would not.

Some (apparently) physically reasonable results could also be
obtained using some solutions of the equations of the pure
gravitational field. For example, the bending of light, the
perihelion precession, the prediction of gravitational radiation
etc. The observations may also ``agree'' with such results.
However, such cases are the ``pathological situations'' of
definite kind.

Just as we cannot conclude the correctness of the newtonian
corpuscular theory of light on the basis of the explanation it
provides for the existence of penumbra on the assumption of ``some
suitable hypothetical force'' acting on the light corpuscles
making them enter the shadow of an object illuminated by light, we
also cannot conclude that the solutions of the equations of the
pure gravitational field provide us physically reasonable
explanations of the observed phenomena.

The meaning of the phrase ``the equations of the pure
gravitational field are ghostly'' is then this above. Clearly, any
such physical explanations of observed phenomena must, therefore,
be based on the case (b) mentioned earlier.

Of course, it may happen in some situations that the mathematical
expression for some phenomena obtained on the basis of case (b) is
``identical'' with that of the case of some solution of the
equations of the pure gravitational field. But, that is for the
case (b) to tell us.

However, this does not mean that such solutions of the pure
gravitational field are some ``good approximation'' to the
corresponding situation of case (b). Just as the explanation of
the formation of penumbra on the basis of some suitable force
acting on light corpuscles is ``ghostly,'' the solutions of the
equations of the pure gravitational field also remain ``ghostly.''

In essence, one must then wait for the case (b) to provide us such
physically acceptable explanations of various observed phenomena.

Such considerations of fundamental difficulties then prompt us to
abandon the case (a) of the pure gravitational field or of the
field-particle dualism in General Relativity, {\em ie}, the notion
of a (curvature) singularity as a mass point or particle. This is
as far as the case (a) of the pure gravitational field is
concerned, then.

(We must therefore, unassumingly, seek the pardon of all those
whose sincere as well as herculean efforts gave us the many
solutions of these highly non-linear equations of the pure
gravitational field. Unfortunately, being ghostly, these equations
cannot, however, further our understanding of the Nature for the
above reasons.)

At this place, we may also note that any (open or concealed)
increasing of the number of dimensions from four does not change
this situation. Clearly, all of the above fundamental problems
associated with a (curvature) singularity will arise in higher
dimensions in exactly the same manner as holds for the case of the
pure gravitational field which we have considered above.

Therefore, unless there are some other specific reasons, it does
not appear compelling to consider higher dimensional situations.

Restricting to four dimensions, we then consider next the
situation of Einstein's ``preliminary equations'' containing
matter in the form of the energy-momentum tensor.

We then recall here the concern expressed by Einstein regarding
these equations. The point of Einstein's concern \cite{ein1} is
that of the comprehension of the concept of an energy-momentum
tensor in the sense of a field theory.

In this connection, we then note that the energy-momentum tensor
is (usually) obtained on the basis of only the particle
considerations. Recall that to obtain the related basic physical
quantities we consider a collection of particles. By defining
various physical quantities (such as, for example, the flux of
particles across a surface), we average these quantities over this
collection of particles. These related (averaged) quantities then
help us define the energy-momentum tensor.

(For a mass point as a singularity of the spacetime, the notion of
its motion is not mathematically available. Therefore, it is clear
that we would not be able to define, for example, the flux of
singularities across a surface. Hence, the energy-momentum tensor
is not obtainable by considering particles as spacetime
singularities.)

Since there cannot be any point-particle, it is not clear what we
mean by a particle in General Relativity. It is only after we have
specified this concept that the question of defining physical
quantities and averaging them over a collection of particles can
be tackled.

Clearly, difficulties arise with the comprehension of the
energy-momentum tensor because it is not clearly specified in the
``field theoretical framework'' of general relativity what exactly
we mean by a ``particle'' (which, of course, cannot be a
singular-particle now). Moreover, it is also not clear as to how
one can unambiguously define the concept of a particle in this
``field-theoretic framework'' of geometric character, except
perhaps as an {\em extended region of energy}.

Then, the energy-momentum tensor of the Einstein field equations
is not a well-defined concept to begin with. Therefore, it is
unclear whether the Einstein field equations can be formulated
without first specifying what we mean by a particle. It is then
equally unclear as to what the solutions of these equations mean
in the absence of a clear formulation of the concept of a particle
(which, now, cannot be a singular-particle).

However, it should be clear by now that only smooth
(singularity-free) spacetime geometries in General Relativity need
to be considered since the energy density should be non-vanishing
everywhere in a spacetime. Many such solutions of Einstein's
makeshift equations, spacetime geometries, are available in the
literature \cite{stdworks}.

The issue therefore arises of extracting physical results out of
these spacetimes, in particular, in the absence of the
availability of the notion of particle. Then, which of these
smooth geometries are to be considered relevant to this Physics
without the field-particle dualism?

It may, however, be noted at this stage that there simply cannot
be any considerations of ``singular-particles'' and ``appropriate
fluid description approximating some collection of particles'' in
all these spacetimes because of the reasons already considered by
us earlier.

It may be stressed once again that such considerations are not
justifiable in any manner. To stress the same thing again, we note
that the concept of energy-momentum tensor is itself {\em not\/}
yet defined by us because we do not have any well-defined notion
of what a particle is in this framework. This is the primary
reason behind the current situation of the above kind.

However, the makeshift equations, the Einstein field equations,
may provide us a useful starting point. This is true provided we
use other physical principles to some advantage. We therefore need
to explore this issue further.

We will, for the time being, postpone these considerations of
smooth spacetimes.

\section{Quantum and its theory} \label{quantumtheory}
Historically, the theory of the quantum owes its origin to the
dualism of a particle and a wave. It can be traced to Newton's
times.

As seen earlier, Newton's geometric considerations led him to
postulate a mass point and his laws (of mechanics and gravitation)
are primarily based on this conception of physical bodies. If
every physical body were to follow these laws, then it must be
possible to treat it using the concept of a mass point.

Therefore, Newton proposed the corpuscular theory for light. That
a ray of light propagates in a straight line, that it is reflected
from a surface (of a mirror) etc.\ are then explainable on the
basis of the corpuscular hypothesis.

However, in Newton's own experiments in optics, it emerged that
the corpuscular hypothesis does not explain, in a natural manner,
all the phenomena displayed by light.

As an example, in the situation of an object illuminated by light,
umbra and penumbra form. As a corpuscle, light is not expected to
penetrate the region forming the shadow of the object. Then, the
existence of penumbra needed some corpuscular explanation.

In another experiment, Newton observed ``Newton's rings''. This
experiment shows that corpuscles of light gather in some regions
forming the (bright) rings while they do not at all gather in
other (dark) regions. This fact also needed some corpuscular
explanation.

On the basis of Newton's laws, it then follows that some force
(acting on light) is causing this behavior of the light
corpuscles. However, this force then acts on only ``some''
corpuscles to pull them to the bright regions but not on some
other light corpuscles (which escape in a straight path behind the
object, for example).

Furthermore, in the case of Newton's rings, there are more than
one bright/dark such regions. Then, the force acting on the light
must be (comparatively) ``larger'' for some corpuscles and
``smaller'' for some other corpuscles.

Such an explanation is definitely thinkable. But, an important
question is now that of the simplicity of this explanation. The
simplicity of a theory has always been the driving impetus behind
the scientific investigations.

It was clear to Newton that this explanation is not appealing. It
requires us to postulate the ``switching on and off'' of the force
whose strength is also different for different corpuscles. What
causes such a behavior of this force? Is there some universal
rational explanation for this?

There did not exist in the newtonian framework any conceivable
explanations of such behavior of the force under consideration
here.

Huygens, on the other hand, considered these phenomena from an
entirely different perspective. His wave theory of light
postulated undulatory behavior for light on the basis of the
continuum postulate. Similar to waves on the surface of ocean, he
imagined light as a wave phenomenon in some (unknown) medium.

Huygens's wave theory of light then provided satisfactory
explanations of the optical phenomena on the basis of constructive
and destructive interference of the waves. The wave theory of
light then gained wide acceptance.

This situation persisted till almost the end of the nineteenth
century. Newton's corpuscular theory for light was then forgotten
or, at least, not considered seriously.

Incorporation of various optical phenomena in Maxwell's
electromagnetism was a climactic stage for the wave theory. It
showed that light consists of oscillations of electric and
magnetic fields transverse to the direction of a propagating light
ray. The polarization of light then received an explanation here.
(We also note that Newton's corpuscular theory had no explanation
whatsoever for this peculiar behavior of light.)

It then came as a real surprise that the corpuscular behavior of
light surfaced in some experiments again, of special significance
is the photo-electric effect discovered by Hertz in 1887.

In Hertz's experiments, a metal is subjected to an incident
radiation. Such a metal ejects electrons if the frequency of
incident radiation is above a certain threshold which depends on
the metal properties. The kinetic energy of ejected electrons does
not, however, depend upon the intensity of incident radiation but
only on the difference of the frequency of incident radiation and
the threshold frequency of the metal.

In the beginning, it was not at all clear as to how one could
explain this Hertzian photo-electric effect in any manner. G
Kirchhoff's investigations into radiation had also provided us
various laws regarding the behavior of heat radiation interacting
with matter. Efforts then began of providing explanations of these
empirical laws on the basis of Maxwell's theory.

No one realized that a stage was slowly getting prepared for a
fundamental crisis with Maxwell's theory of electromagnetism in
this era of hectic experimental activities.

In 1900, Max Planck's remarkable intuition in dealing with deep
physical problems suddenly brought into focus the seriousness of
this crisis with classical theories.

These investigations are all the more remarkable in that they were
based primarily on only the theories regarding the behavior of
heat radiation interacting with matter.

In what follows, we (partly) adopt Einstein's (method of)
exposition \cite{ein1} of these theoretical developments, because
his exposition remarkably clearly describes the (theoretical)
nature of this fundamental crisis.

Then, let us first note that Kirchhoff had concluded, on
thermodynamical grounds, that the energy density and the spectral
composition of radiation in a cavity (Hohlraum) with impenetrable
walls of absolute temperature $T$ is independent of the material
of the walls. That is to say, the monochromatic density,
$\varrho$, of radiation is some {\em universal\/} function of the
frequency $\nu$ of radiation and of the absolute temperature $T$.

Thus arose the problem of determining this universal function
$\varrho(\nu,T)$.

As per Maxwell's theory, radiation exerts pressure on the walls of
the cavity and this pressure is determined by the total energy
density. From this, Boltzmann then concluded that the entire
energy density of radiation, $\int \varrho\,d\nu$, is proportional
to $T^4$ and thereby provided a theoretical explanation of
Stefan's empirical law on the basis of Maxwell's theory. W Wien
then used ingenious thermodynamical arguments, also using
Maxwell's theory, to deduce that \[\varrho \sim
\nu^3\,f\left(\frac{\nu}{T}\right)\] Clearly, $f$ is a universal
function of only one variable $\nu/T$ and its theoretical
determination is of definite importance.

Basing his faith on the empirical form of the function $f$, Planck
firstly succeeded in reaching the following form for $\varrho$ \[
\varrho = \frac{8\pi\,h\,\nu^3}{c^3}\,\frac{1}{e^{h\nu/kT}-1}\]
whereby he had two universal constants $k$ and $h$ in the
expression for the monochromatic energy density $\varrho$ of the
radiation.

If this formula were correct, it permitted the calculation of the
average energy $E$ of an oscillator interacting with the radiation
in the cavity: \[ E=\frac{h\,\nu}{e^{h\nu/kT}-1} \] For fixed
$\nu$ but high temperature, this gives $E=kT$ - an expression
obtainable from the kinetic theory of gases. From this theory, we
have $E=(R/N)T$ where $R$ is the gas constant and $N$ is the
famous Avogadro's number.

Then, $N=R/k$ and its numerical value agreed reasonably well with
that of the kinetic theory of gases. Therefore, Planck's
investigations were in agreement with the size of the atom since
Avogadro's number tells us about it.

Using Boltzmann's (and Gibbs's) entropy methods, Planck then
partitioned the total energy into a large but finite number of
identical bins of size $\epsilon$ and asked in how many different
ways can $\epsilon$ be divided among the oscillators. This number
would then furnish the entropy and, hence, the temperature of the
cavity-radiation system.

As is well known, Planck obtained the aforementioned radiation
formula if $\epsilon=h\,\nu$. His expression then yields correctly
the Rayleigh-Jeans, the Stefan-Boltzmann, the Wien laws.

However, Planck's reasoning camouflaged the fact that his
derivation demands that energy can be emitted and absorbed by an
oscillator only in {\em quanta\/} of energy $\epsilon=h\,\nu$.

Planck's formula implies that the energy of any arbitrary
mechanical system capable of oscillations can be transferred only
in ``packets'' or these quanta. The same is also the situation
with the energy of radiation. This contradicts fundamentally the
laws of Newton's mechanics as well as those of Maxwell's
electrodynamics.

Then, we note that Planck's formula is {\em compatible\/} with
Maxwell's electrodynamics, although it is not a necessary
consequence of its equations. Therefore, the contradiction with
Newton's mechanics is (more) fundamental here than that with the
electrodynamics.

This was clear to Einstein soon after the appearance of Planck's
fundamental work. Although he had no ideas on what framework(s)
should substitute Newton's and Maxwell's theories, Einstein's
intuition nonetheless permitted him to apply Planck's formula to
explain the photo-electric effect in a remarkably simple fashion.

If the implication of Planck's reasoning were correct, then an
electron can absorb the energy of radiation only in quanta. An
electron (bound to an atom) could then be set free (ejected) on
absorption of energy of the incident radiation only if the
frequency of incident radiation were larger than certain minimum
corresponding to the binding energy of the electron. There is a
threshold then for the frequency of incident radiation below which
electron ejection does not occur in this hertzian photo-electric
effect.

N Bohr's remarkable insights developed the quantum theoretical
explanations for the empirical laws of atomic spectra. Sommerfeld
then also included special relativistic considerations in Bohr's
theory of the atom. These inputs were of radical theoretical
nature indeed.

Einstein followed with interest these works and, in his own turn,
revealed the deep connection between Planck's formula and Bohr's
law of frequencies, thereby introducing the probabilities of
quantum transitions of atomic systems. This is where the Einstein
coefficients for induced and spontaneous transitions made their
inroad.

This is the return, in a sense, of Newton's corpuscle - a particle
of light and, hence, of the wave-particle dualism (for light). We
will however not go into details of other developments here.
Instead, we turn to the next important step that was taken for the
concept of a quantum.

Next, L de Broglie proposed that if light, primarily an
electromagnetic wave in Maxwell's theory, displays particle-like
phenomena, then particles (of Newton's type) should display
wave-like phenomena in order that the basic symmetry of wave
versus particle is maintained.

This radical proposition, of course, needed experimental
confirmation and it was soon obtained in diffraction experiments.

Einstein then generalized S N Bose's statistical methods (for
light particles) to particles indistinguishable from each other.
This goes in the name of the Bose-Einstein statistics and the
particles obeying these statistical methods are now called as the
Bosons.

The fundamental difference between the statistical properties of
like and unlike particles is intimately connected with the
circumstance that, due to Heisenberg's indeterminacy relations,
the possibility of distinguishing between like particles, with the
help of the continuity of their motion in space and time, is
getting lost. Pauli's analytical mind grasped this fundamental
issue immediately and that is what led him to (Pauli's) exclusion
principle for electrons.

Fermi and Dirac then showed that electrons, in particular, follow
a different statistical method since they obey Pauli's exclusion
principle. This goes in the name of the Fermi-Dirac statistics and
the particles obeying these statistical methods are now called as
the Fermions.

It is still a deep mystery as to whether the particles of Nature
are only of these two types, Bosons and Fermions. If yes, why.

The need for replacement(s) of Newton's and of Maxwell's theories
became urgent.

E Schr\"{o}dinger then formulated the Wave Mechanics and, almost
simultaneously, W Heisenberg formulated the Matrix Mechanics for
the quantum. These works provided the bridge between the particle
and wave conceptions. M Born then provided the ``probability
interpretation'' of Schr\"{o}dinger's Wave Mechanics and Bohr
supported it with his complementarity arguments.

The phenomenon of barrier penetration (tunnelling) was also
discovered in such considerations. Special relativistic
considerations were taken up by P A M Dirac and others. These
provided us the concept of particle and anti-particle pair. The
implications of such considerations were ``confirmed'' in numerous
experiments. Today, these considerations are the basis of numerous
technological equipments and also of their theories.

The method of {\em second or field quantization\/} was
subsequently developed. The primary thesis of this method is that
we should be able to {\em count\/} the number of quanta, if these
are really the packets of energy. It is this (second quantization)
method that has come to be recognized as the {\em genuine\/}
theory of the quantum.

The theory of the quantum that developed as a result was
mathematically satisfactory but extremely counter-intuitive. We
then note here that many vigorous discussions and efforts were
needed to reconcile the results, in particular, those concerning
Heisenberg's indeterminacy relations, with the physical intuition.
Rather than entering these historical details, we then refer to
various excellent essays, in particular, Bohr's article and
Einstein's Reply to Criticisms in \cite{ein1}.

\section{Criticism of the Theory of the Quantum}
\label{criticismofquantum}

Let us first recollect that Einstein, in spite of his initial
resistance as reported in Bohr's essay in \cite{ein1}, finally
agreed \cite{ein2} to the correctness of Heisenberg's
indeterminacy relations. However, he differed from most other of
his contemporary physicists on the issue of probability being the
only basis of understanding the entire physical world.

In Einstein's own words \cite{ein2} (my curly brackets): {\em ...
On the strength of the successes of this theory they \{Born,
Pauli, Heitler, Bohr, and Margenau\} consider it proved that a
theoretically complete description of a system can, in essence,
involve only statistical assertions concerning the measurable
quantities of this system. They are apparently all of the opinion
that Heisenberg's indeterminacy relation (the correctness of which
is, from my own point of view, rightfully regarded as finally
demonstrated) is essentially prejudicial in favor of the character
of all thinkable reasonable physical theories in the mentioned
sense. ...}

He further hastened to add to the above that: {\em ... I am, in
fact, firmly convinced that the essentially statistical character
of contemporary quantum theory is solely to be ascribed to the
fact that this [theory] operates with an incomplete description of
physical systems.}

Still, it is undoubtable that the self-consistent formalism of the
Quantum Theory provides us the theoretical explanations of variety
of experimental results. In fact, the amount of experimental data
it supports is so enormous, so unparalleled in the history of
Physics, that there must be some element of the finality in its
formalism.

Einstein did not want to leave behind any doubts that he did not
recognize the importance of the contributions from the Quantum
Theory. So, he added further: {\em ... This theory is until now
the only one which unites the corpuscular and undulatory dual
character of matter in a logically satisfactory fashion; and the
(testable) relations, which are contained in it, are, within the
natural limits fixed by the indeterminacy relation, {\em
complete}. The formal relations which are given in this theory -
i.e., its entire mathematical formalism - will probably have to be
contained, in the form of logical inference, in every useful
future theory.}

Then, Einstein went on to tell us what exactly it is that does not
satisfy him with this Theory of the Quantum: {\it What does not
satisfy me in that theory, from the standpoint of principle, is
its attitude towards that which appears to me to be the
programmatic aim of all of physics: the complete description of
any (individual) real situation (as it supposedly exists
irrespective of any act of observation or substantiation).}

In what follow we recall (although not in a verbatim manner)
Einstein's arguments in support of his above statement
\cite{ein1}.

Let us then consider a radioactive atom as a physical system. For
practical purposes, we can consider that it is located exactly at
a point of the coordinate system. We may also neglect the motion
of the residual atom after its radioactive disintegration process
in which a (comparatively light) particle is emitted by the atom.
Then, following Gamow's theory, we may replace the rest of the
atom by a potential barrier which surrounds the particle to be
emitted. The radioactive disintegration process is then the
``tunnelling'' of the particle out of this potential barrier.

We then solve Schr\"{o}dinger's equation and obtain
Schr\"{o}dinger's $\Psi$-function which is initially nonzero only
inside the potential barrier, but which, with time, becomes
non-vanishing outside the barrier. Essentially, the
$\Psi$-function yields the probability of finding the initially
``trapped'' particle to be outside of the trapping barrier, at
some later time, in some specific portion of the space outside of
that potential barrier or the atom.

However, this does not imply any assertion of the time-instant of
the disintegration of the radioactive atom. That is, no {\em
observable\/} exists in the quantum theory for this time-instant.

Einstein then raised the question: Can this theoretical
description be taken as the {\em complete\/} description of the
disintegration of a single individual atom? The immediately
plausible answer is: No, he wrote. For we are inclined to assume
the existence of an instant of the disintegration and such a
definite value of the time-instant is not implied by the
$\Psi$-function.

Einstein then answered for a quantum theorist: This alleged
difficulty arises from the fact that one postulates something not
observable as ``real.'' That is to say, this difficulty arises
because one is assuming the (reality of) time-instant of the
disintegration of an individual atom that is not an {\em
observable\/} of its (quantum) theory.

Einstein then phrased the question as: Is it, within the framework
of our theoretical total construction, reasonable to assume the
existence of a definite time-instant of the disintegration of an
individual atom?

Then, if one takes the viewpoint that the description in terms of
a $\Psi$-function refers only to an ideal systematic totality
(ensemble) but not to an individual system, then one may assume
the existence of the time-instant of disintegration of an
individual atom.

But, if one represents the assumption that the description in
terms of the $\Psi$-function is a {\em complete\/} description of
an individual system, then one must reject the existence of the
time-instant of an individual atom and can justifiably point to
the fact that a determination of the exact time-instant of
disintegration is not possible for an individual atom. Any such
attempt would mean disturbances (of the atom) of such nature which
would then destroy the very phenomenon whose time-instant we are
trying to determine.

Now, following Schr\"{o}dinger's reasoning, one may construct a
contraption which kills a cat (macroscopic object) sitting close
to the radioactive atom only if the decay particle is emitted by
the atom. One may then ask: Is the cat alive or dead at some later
instant of time? Here, one would expect to get an answer with
certainty since a beam of torchlight falling on the cat cannot be
expected to disturb (kill it, if alive) its state. But, the theory
of the quantum can only tell us the probability of the cat being
alive or dead.

Then, to begin with we have the $\Psi$-function for an alive cat.
With time, the $\Psi$-function is a {\em superposition\/} of two
components, one for the alive cat and one for the dead cat. It is
only when an observer makes an observation (of the cat) that the
$\Psi$-function reduces or {\em collapses\/} to one of (these two
of) its components. This collapse of the $\Psi$-function
represents the interference of an observer with the system under
observation.

Surely, the quantum theory as represented by Schr\"{o}dinger's
$\Psi$-function is self-consistent. Now, if this probabilistic
theory of the quantum is of universal character, that is to say,
if the basic laws of nature are intrinsically probabilistic in
character, then this formalism applies to microscopic as well as
macroscopic systems.

Some of the macroscopic systems constitute experimental equipments
and, hence, their behavior is also probabilistic in character
then.  By associating Schr\"{o}dinger's $\Psi$-function with {\em
every\/} physical system, the theory of the quantum treats then
measuring apparatuses and all other things on an equal footing in
its formalism.

But, a question can then be asked as to when, within this
formalism of the quantum theory, can we say that the ``measuring
apparatus'' has made its observation.

It is evident that this question is related to the collapse of the
$\Psi$-function because there is no another way of answering it
within the realm of the quantum theory. Therefore, we have to say
that the system makes an observation only when the collapse of the
$\Psi$-function takes place to one of its various components, to
that which uniquely corresponds to the value of the observable
measured by the macroscopic apparatus.

Then, we can consider an apparatus which has been ``suitably
prepared'' to measure a specific physical observable of a certain
physical system. But, let us not interfere with this apparatus in
any manner whatsoever and let us even not attempt to `look'' at
the reading of the apparatus. But, now, let us ask a question as
to whether the apparatus has ``performed'' the measurement of that
quantity it were prepared to measure by ``prearrangement of some
special kind.''

For example, we prepare the apparatus to locate the position of an
electron in Heisenberg's microscope and place a charge-coupled
device (CCD) at the eyepiece so that a photon, reflected after the
collision with an electron, makes a mark on it. This action in
Heisenberg's microscope takes place, but, we choose not to look at
that mark made by that photon on the CCD plate.

Then, the question is: whether this apparatus has performed the
measurement of the position of an electron, particularly when we
have not looked at the CCD plate.

Evidently, if the formalism of the quantum theory were to tell us
that the above apparatus has ``never'' performed any observation,
then the result of the observation is {\em not\/} known to any
{\em observer}, that is to say, the collapse of the
$\Psi$-function has not occurred for any observer. This is
indicative of the importance of an observer in the framework of
the quantum theory.

On the other hand, if we say that an observation has been
performed in the above prearrangement with Heisenberg's
microscope, then the $\Psi$-function will always be in some
collapsed state because the phenomenon occurring in Heisenberg's
microscope (for that matter, any other apparatus) occurs
everywhere in the space. This then results in an obvious absurdity
with the entire formalism of the quantum theory.

Therefore, we are forced to assume here that no observation is
made in the concerned prearrangement with Heisenberg's microscope
or with any other prearranged apparatus. An observer is therefore
of definite importance in this interpretation of the quantum
theory.

Now, an observer is also made up of the same thing (matter) that
an apparatus is made up of. Hence, the above situation also
applies to an observer (as a measuring apparatus, may be of
special kind). It is therefore not specified by the quantum theory
as to what it means by an observer and, hence, what precisely
constitutes an act of observation within its formalism.

Surely, although complicated constructs, as is an apparatus or an
observer, are not explicitly describable, one can ascribe a
corresponding total $\Psi$-function to them, thereby rendering the
phenomena of the macroscopic world also probabilistic in
character. Surely, the measuring apparatuses are then {\em at
par\/} with everything the quantum theory treats. However, some
questions of serious scientific concern then arise.

Clearly, an observer is  then given an exceptional importance in
the theory of the quantum but this theory does not tell us what
constitutes an observer and how does an observer act to collapse
the $\Psi$-function on having made an observation of some
(quantum) system.

Then, the issue arises as to where, in some appropriate sense,
does the collapse (of the $\Psi$-function) occur? How is it to be
described in some understandable language? Many such questions can
be raised.

In response to such questions, we may, for example, then propose
that the collapse occurs in the mind of a conscious observer. Such
an approach but transgresses the obvious limits of the scientific
inquiry and, hence, invites the corresponding criticism. This is
then the problem of the collapse of the $\Psi$-function.

To this date, there do not appear to exist any genuinely
satisfactory resolutions \footnote{See also Section
\ref{totalfield}.} of such paradoxes of the quantum theory.

Schr\"{o}dinger's Cat Paradox and other paradoxes therefore
highlight such problematic issues of the quantum theory if this
theory were assumed to be universal of character.

This is surely then a problematic issue for the quantum theory
(description using Schr\"{o}dinger's $\Psi$-function) if it is to
be of universal character, that is, if its laws are to be
universally applicable to all the physical systems.

Hence, from the point of view of a {\em complete\/} theory forming
the basis of the totality of physical phenomena, in a sense
similar to that of Newton's theory (supposedly) forming the basis
for the totality of physical phenomena, we therefore note the
following important lacunae with the theory of the quantum
described above.

Firstly, the theory of the quantum, as it stands even today, does
not incorporate the explanations needed for the equality of the
inertial and the gravitational mass of a physical body, which is
to be taken as an experimental result.

Secondly, this theory separates artificially ``an observer'' (as
definable in the theory of the quantum) and all other things that
it treats since there is no consistent demarkation line between
what constitutes ``an act of observation'' in this theoretical
framework. The considered example of the ``prearranged''
Heisenberg's microscope highlights precisely this issue.

In this connection, it cannot be forgotten that the measuring
apparatuses are also made up of the same things that the formalism
of the theory is supposed to treat. Therefore, their treatment in
a theory must be at par with everything else that the theory
intends to treat, this is if the theory claims universality of its
formalism. Then, we would not expect to find any serious
paradoxical situations raised in the theory.

Then, we note that, in relation to the standard viewpoint
regarding the probabilistic interpretation of the laws of the
quantum theory, serious paradoxical situations have been
constructed and the resolutions of these paradoxical situations
have not been achieved. Clearly, the emphasis of these paradoxes
is on whether the quantum theory is of universal character,
whether the basic laws of nature are probabilistic of character.

If we assume that the quantum theory is of universal character,
then we end up with the paradoxical situations of serious
character, with the quantum theory not offering us any
satisfactory resolutions of these paradoxes.

On the other hand, if we assume that it is not of universal
character, then we must search for an alternative formulation
which evidently cannot be based on the (probabilistic) formalism
of the quantum theory. That is to say, the laws as are applicable
to all the physical systems cannot then be probabilistic of
character. Question is then of such an alternative formulation for
the description of the totality of physical phenomena.

It may therefore be noted here that various paradoxes of the
quantum theory then indicate that \cite{ein1}: {\em The attempt to
conceive the quantum-theoretical description as the complete
description of the individual systems leads to unnatural
theoretical interpretations, which become immediately unnecessary
if one accepts the interpretation that the description refers to
ensembles of systems and not to individual systems.}

Einstein \cite{ein1} then continued: {\em There exists, however, a
simple psychological reason for the fact that this most nearly
obvious interpretation is being shunned. For if the statistical
quantum theory does not pretend to describe the individual system
(and its development in time) completely, it appears unavoidable
to look elsewhere for a complete description of the individual
system; in doing so it would be clear from the very beginning that
the elements of such a description are not contained within the
conceptual scheme of the statistical quantum theory. With this one
would admit that, in principle, this scheme could not serve as the
basis of theoretical physics. Assuming the success of efforts to
accomplish a complete physical description, the statistical
quantum theory would, within the framework of future physics, take
an approximately analogous position to the statistical mechanics
within the framework of classical mechanics. ...}

Perhaps, there is an element of truth in these (last quoted)
Einstein's opinions. Next, we shall see that a general
relativistic theory of the total field does offer such a
possibility.

\section{A Theory of the Total Field} \label{totalfield}

Let us then return to the considerations of the smooth spacetime
geometries in general relativity that are obtainable on the basis
of Einstein's makeshift field equations.

As noted before, in this case, to extract physical results in the
absence of any conception of a particle, which cannot be a
point-particle, we need to employ other physical principles to
some advantage. That this is indeed possible is what is the
subject of the present section.

It is of course not very clear to begin with as to which physical
principles to use to extract meaningful results in this situation.
However, we must look for some hints here.

To begin with, as we have seen before, considerations of the
``pure or vacuum gravitational field'' lead us to an internal
inconsistency in the formalism. Then, we note that the energy
density of ``matter'' must be non-vanishing at every spatial
location in any such (smooth) spacetime.

Also, the volume-form, in Cartan's sense, is well-defined at every
location in this (smooth) spacetime. Consequently, just as it was
permissible to attribute the concept of inertia to a point of
space in the newtonian situation, we should also be able to define
the concept of inertia for (every) spatial location of the
(smooth) spacetime, of course, in only some non-singular sense.

Then, a well-behaved (Cartan's) volume-form can be expected to
allow us appropriate definitions of non-vanishing
``gravitational,'' ``inertial'' and also ``total'' mass for
(every) point of the space in such a spacetime. (See later.)

Next, the case (b) under consideration is, clearly, that of the
general field in which quantities corresponding somehow to an
electric field occur as well. Therefore, a point of the space in
such a spacetime can also be prescribed an electric charge as a
source attribute of a physical body in a manner similar to the
case of the total mass attribute.

This must also be the case with all the permissible source
attributes of a physical body since the same procedure can be
expected to work for each of such source attributes.

Thus, it must be possible to incorporate all the source attributes
of a physical body in it. Then, in this spacetime, physical bodies
are concentrated {\em total\/} energy \cite{ein1} and would {\em
everywhere in space\/} be describable as singularity-free. We may
also look at a physical body in the newtonian, now non-singular,
sense of a point particle possessing the source attributes as
outlined above.

Next, any local motion of a physical body can, clearly, be a
change in the local energy distribution in this spacetime.

However, the global properties of the spacetime must not change
with any ``local'' motions of a physical body in it. Otherwise,
infinite speeds of communication exist in the spacetime. This is,
physically speaking, undesirable.

A spacetime for which all the spatial properties are arbitrary has
these desired properties. (See below.) The required spacetime is
given by \cite{smw00}: \setcounter{equation}{0}
\beq ds^2 = &-& P^2Q^2R^2dt^2 +{P'}^2Q^2R^2B^2 dx^2 \n \\
&+& P^2\bar{Q}^2 R^2C^2dy^2 +P^2Q^2\tilde{R}^2D^2\,dz^2
\label{ghsp} \eeq  where $P\equiv P(x)$, $Q\equiv Q(y)$, $R\equiv
R(z)$, $B\equiv B(t)$, $C\equiv C(t)$, $D\equiv D(t)$. We also use
$P'=dP/dx$, $\bar{Q}=dQ/dy$ and $\tilde{R}=dR/dz$.

Now, consider an energy-momentum tensor (with heat flux) for the
fluid in the spacetime and Einstein's (makeshift) field equations
using it. The energy-density in the spacetime of (\ref{ghsp}) then
varies as $\rho \propto {1}/{P^2Q^2R^2}$ and is seen \cite{smw00}
to be {\em arbitrary\/} because the field equations do not
determine the spatial functions $P$, $Q$, $R$.

But, we must remember that we do not know whether the
energy-momentum tensor is {\em abinitio\/} definable one for this
spacetime under consideration. (However, see later.) Thus, it is
not truly justifiable to use it to obtain the Einstein makeshift
field equations for (\ref{ghsp}).

However, what we realize here is that it is also the 3-space, any
constant-time section of (\ref{ghsp}), that is endowed with
various properties such as curvature, pseudo-metric nature etc.

Let us recall here that the characteristic of Newtonian Physics is
that it ascribes independent and real existence to space and time
as well as to matter. In this newtonian formulation, space and
time play a dual role.

Firstly, they play the role of a background for things happening
physically. Secondly, they also provide us the inertial systems
which happen to be advantageous to describe the law of inertia.
Therefore, if matter were to be somehow removed completely, the
space and time of the newtonian framework would ``remain'' behind.

Let us also recall Descartes's opposition to consider space as
independent of material objects \cite{ein-pop}: space is identical
with extension, but extension is connected with physical bodies;
thus there should be no space without physical bodies and hence no
empty space \footnote{There certainly are (philosophical)
weaknesses of such an argument. However, we shall not enter the
relevant issues here. It suffices for our purpose here to say that
``the space must be indistinguishable from physical bodies.''}.

We then notice that this expectation is indeed true of the spatial
sections of (\ref{ghsp}). Physical bodies are ``extended regions
of space'' in it.

The issue, however, remains of incorporating time in the framework
of (\ref{ghsp}), in particular, in the absence of a well-defined
concept of a particle. Here, Einstein's makeshift equations
provide us the required clue to this issue.

The makeshift equations essentially provide the laws for the
temporal evolution of physical bodies in (\ref{ghsp}). Then, the
makeshift equations can, in turn, be determined on the basis of
the temporal evolution of physical bodies in (\ref{ghsp}).

It may now be stressed that the makeshift equations are based on
the as-yet undefined concept of the energy-momentum tensor for
matter fields, undefined because it is not clear as to how to
define the concept of a physical particle in this theoretical
framework as yet.

It is therefore logically compelling to investigate whether the
second alternative, that of determining the makeshift equations on
the basis of the temporal evolution of physical bodies, extended
space in (\ref{ghsp}), is the realizable one.

In fact, this last approach is really the (logically) appropriate
one since the temporal evolution of ``physical bodies'' in
(\ref{ghsp}) is a mathematically well-definable concept (dynamical
systems) while the makeshift equations are not.

We therefore abandon \cite{smw01} the four-dimensional spacetime
in favor of a three-dimensional pseudo-Riemannian manifold
admitting a pseudo-metric, called the Einstein pseudo-metric,
given as:
\beq d\ell^2&=& {P'}^2Q^2R^2\, dx^2 \n \\
&\phantom{m}&\hspace{.1in} +\;P^2\bar{Q}^2 R^2\, dy^2 \n \\
&\phantom{m}&\hspace{.3in} +\;P^2Q^2\tilde{R}^2 \,dz^2
\label{3d-metric-gen} \eeq where, as before, $P\equiv P(x)$,
$Q\equiv Q(y)$, $R\equiv R(z)$ and $P'=dP/dx$, $\bar{Q}=dQ/dy$,
$\tilde{R}=dR/dz$. We denote the space of (\ref{3d-metric-gen}) by
the symbol $\mathbb{B}$. The three spatial functions $P$, $Q$, $R$
are {\em initial data\/} for the space $\mathbb{B}$. The vanishing
of any of these spatial functions is a {\em curvature
singularity}, and constancy (over a range) is a {\em degeneracy\/}
of (\ref{3d-metric-gen}).

A particular choice of functions, say, $P_o$, $Q_o$, $R_o$ is a
specific spatial distribution of energy in the space of
(\ref{3d-metric-gen}). As some ``concentrated'' energy ``moves''
in the space, we have the original set of functions changing to
the ``new'' set of corresponding functions, say, $P_1$, $Q_1$,
$R_1$.

Then, ``motion'' as described above is, basically, a {\em change
of one set of initial data\/} to {\em another set of initial
data\/} with ``time''.

Clearly, we are considering the isometries of
(\ref{3d-metric-gen}) while considering ``motion'' of this kind.
Then, we will remain within the group of the isometries of
(\ref{3d-metric-gen}) by restricting to the triplets of {\em
nowhere-vanishing\/} functions $P$, $Q$, $R$. We also do not
consider any degenerate situations for (\ref{3d-metric-gen}).

If we denote by $d$, a metric function canonically \cite{kdjoshi}
obtainable \footnote{We define an equivalence relation ``$\sim$''
such that $x\sim y$ iff $\ell(x,y)=0$ where $\ell$ is the
pseudo-metric distance defined on the space $X$. Denote by $Y$ the
set of all equivalence classes of $X$ under the equivalence
relation $\sim$. If $A,B \in Y$ are two equivalence classes, then
let $e(A,B)=\ell(x,y)$ where $x\in A$ and $y\in B$. The (metric)
function $e$ on $Y$ is the canonical distance.} from the
pseudo-metric (\ref{3d-metric-gen}), then the space $(\mathbb{B},
d)$ is an uncountable, separable, complete metric space. If
$\Gamma$ denotes the metric topology induced by $d$ on
$\mathbb{B}$, then $(\mathbb{B}, \Gamma)$ is a Polish topological
space. Further, we also obtain a Standard Borel Space
$(\mathbb{B},\mathcal{B})$ where $\mathcal{B}$ denotes the Borel
$\sigma$-algebra of the subsets of $\mathbb{B}$, the smallest one
containing all the open subsets of $(\mathbb{B},\Gamma)$
\cite{trim6}.

But, the Einstein pseudo-metric (\ref{3d-metric-gen}) is a metric
function on certain ``open'' sets, to be called the P-sets, of its
Polish topology $\Gamma$. A P-set of $(\mathbb{B},d)$ is therefore
never a singleton subset, $\{ \{x\}:x\in \mathbb{B}\}$, of the
space $\mathbb{B}$. Note also that every open set of $(\mathbb{B},
\Gamma)$ is {\em not\/} a P-set of $(\mathbb{B}, d)$.

Now, the differential of the volume-measure on $\mathbb{B}$,
defined by (\ref{3d-metric-gen}), is \be d\mu \;=\;
P^2Q^2R^2\,\left( \frac{dP}{dx}\frac{dQ}{dy}
\frac{dR}{dz}\right)\;dx\,dy\,dz \label{volume1} \ee This
differential of the volume-measure vanishes when any of the
derivatives, of $P$, $Q$, $R$ with respect to their arguments,
vanishes. (Functions $P$, $Q$, $R$ are non-vanishing over
$\mathbb{B}$.)

A P-set of the space $\mathbb{B}$ is then also thinkable as the
interior of a region of $\mathbb{B}$ for which the differential of
the volume-measure, (\ref{volume1}), is vanishing on its boundary
while it being non-vanishing at any of its interior points.

Any two P-sets, $P_i$ and $P_j$, $i,j\;\in\;\mathbb{N}$,
$i\,\neq\,j$, are, consequently, {\em pairwise disjoint sets\/} of
$\mathbb{B}$. Also, each P-set is, in its own right, an
uncountable, complete, separable, metric space.

A P-set is the mathematically simplest form of ``localized'' total
energy in the space $\mathbb{B}$. This suggests that we should use
set-theoretic concepts for it. One such concept is of a suitable
(Lebesgue) measure definable on sets.

We then recall that the Galilean concept of the (inertial) mass of
a physical body is that of the measure of its inertia. Therefore,
some appropriate measure definable for a P-set is the property of
inertia of a physical body, a P-set in question. So also should be
the case with the gravitational mass of a physical body. Such
should also be the case with other relevant properties of physical
bodies, for example, its electric charge.

Thus, we associate with every attribute of a {\em physical body},
a suitable class of (Lebesgue) measures on such P-sets. Therefore,
a P-set is a {\em physical particle}, always an {\em extended
body}, since a P-set cannot be a singleton set of
$(\mathbb{B},d)$.

Now, in the absence of the field-particle dualism, the field and
the source-particle are indistinguishable. The source-properties
are then also the field-properties. Thus, a P-set (the total
field) and the measures on P-sets (sources) are, then, are
amalgamated into one thing here, {\em ie}, are {\em
indistinguishable\/} from each other, in a sense.

Therefore, various source-properties (measures) {\em change\/}
only when the field (P-set) changes. This union of the field and
the source-properties is then clearly perceptible here.

Moreover, a given measure can be (Haar) integrated over the
underlying P-set in question. The integration procedure is always
a well-defined one here as a P-set, being an ``open'' subset of a
continuum, is a non-empty locally compact Hausdorff group in an
obvious sense.

The (Haar) integral provides then an ``averaged quantity
characteristic of a P-set'' under question. Of course, this is a
property of the entire P-set under consideration.

For example, let us define an almost-everywhere finite-valued
positive-definite measurable function, $\rho$, on $(\mathbb{B},
\mathcal{B})$ and call it the energy density. Integrating it over
the volume of a P-set, the resultant quantity can be called a {\em
total mass}, $m_{_T}$, of that P-set under consideration. The
total mass, $m_{_T}$, is a property of that entire P-set and,
hence, of {\em every point\/} of that P-set.

(Clearly, to define the notions of ``gravitational mass'' and of
``inertial mass'' of a P-set, we need to consider the ``motion''
of a P-set and also an appropriate notion of the ``force'' acting
on that P-set. Since we are yet to define any of these associated
notions, we call the integrated energy density as, simply, the
total mass of the P-set.)

We note that every point of the P-set is then thinkable as having
these averaged properties of the P-set and, in this precise
mathematically non-singular sense, is thinkable as a
point-particle possessing those averaged properties. It is in this
non-singular sense that we can recover the notion of a point
particle in the present framework. That this is indeed permissible
in a mathematically precise sense is then an indication of the
internal consistency of the present approach.

Clearly, the ``location'' of the mass $m_{_T}$ will be {\em
intrinsically indeterminate\/} over the {\em size\/} of that P-set
because the averaged property is also the property of every point
of the set under consideration. We may then associate a Dirac
$\delta$-distribution with the mass $m_{_T}$ over that P-set.

Thus, ``averaging a given measure'' over any P-set and associating
a Dirac $\delta$-distribution with that averaged measure, an {\em
intrinsic indeterminacy\/} of location over the size of that P-set
is obtained for that averaged measure.

Now, any two P-sets of the {\em same cardinality}, belonging
either to the same metric-space $(\mathbb{B}, d)$ or to two
different metric-spaces $(\mathbb{B}, d_1)$ and $(\mathbb{B},
d_2)$, are Borel-isomorphic \cite{krp}. Then, copies (P-sets) of a
physical particle are indistinguishable from each other except for
their spatial locations.

Let us reserve the word {\em particle} for a P-set since it is the
simplest form of ``localized'' energy in the present framework.

Then, in a precise mathematical sense \cite{kdjoshi}, sets can be
touching and that describes our intuitive notion of touching
physical bodies. Of course, the corresponding point particles are
then ``touching'' within the limits of the sizes of the
corresponding P-sets. (The size of a P-set then acts in the manner
of the de Broglie wavelength for a point particle, a point of the
P-set.)

Further, if a P-set splits into two or more P-sets, we have the
process of {\em creation of particles\/} since the measures are
now definable individually over the split parts, two or more
P-sets. On the other hand, if two or more P-sets unite to become a
single P-set, we have the process of {\em annihilation of
particles\/} since the measures are now definable over a single
P-set.

Clearly, the {\em laws of creation and annihilation of
particles\/} will require of us to specify the corresponding
transformations causing the splitting and the merger of the
P-sets.

Now, we call as {\em an object\/} a region of $\mathbb{B}$ bounded
by the vanishing of (\ref{volume1}) but containing interior points
for which it vanishes (so such a region is not a P-set). Such a
region of $\mathbb{B}$ is then a collection of P-sets. But, a
P-set is a particle. Therefore, an {\em object\/} is a {\em
collection\/} of particles.

Objects may also unite to become a single object or an object may
also split into two or more than two objects under transformations
of P-sets. We may then also think of the corresponding laws for
these processes involving objects.

Obviously, various concepts such as the density of particles, a
flux of particles across some surface etc.\ are then well
definable in terms of the transformations of P-sets and the
effects of these transformations on the measures definable over
the P-sets under consideration.

Then, such ``averaging procedures'' are well-defined over any
collection of P-sets and, also, of objects. Thus, we may, in a
mathematically meaningful way, define a suitable ``energy-momentum
tensor'' \footnote{This is a field-theoretic comprehension, in a
definite sense, of the energy-momentum tensor.} and some relation
between the averaged quantities, an ``equation of state'' defining
appropriately the ``state of the fluid matter'' under
consideration.

(Such conceptions require however the notion of transformations of
P-sets and objects. Moreover, this averaging is a ``sum total'' of
the effects of various such transformations of P-sets and objects
and, hence, will require corresponding mathematical machinery.
This is, then, the premise of the ergodic theory. Recall that
$(\mathbb{B},\mathcal{B})$ is a Standard Borel Space.)

Einstein's makeshift field equations are then definable in the
sense (only) of these averages. Therefore, Einstein's makeshift
equations are ``obtainable'' on the basis of the temporal
evolution of points of the space $\mathbb{B}$, physical particles
as elements of the 3-space of $\mathbb{B}$. This is also the sense
in which Descartes's conceptions are then realizable in the
present formalism. However, we will not pursue this obvious issue
of details here.

Now, we can ``count'' P-sets and, also, objects. This precise
mathematical notion of countability of P-sets and objects here
then agrees well with our very general experience that arbitrary
physical objects (chairs, stones, persons etc.) are ``countable''
in Nature in an obvious sense.

Moreover, the metric of $(\mathbb{B}, d)$ allows us the precise
definition of the sizes of P-sets and objects. Then, given an
object of specific size, we may use it as a {\em measuring rod\/}
to measure ``distance'' between two other objects.

A measurable, one-one map of $\mathbb{B}$ onto itself is a Borel
automorphism. Now, the Borel automorphisms of
$(\mathbb{B},\mathcal{B})$, forming a group, are natural for us to
consider here.

Let a Borel automorphism $\phi$ of $\mathbb{B}$ act to take a
point $x_1\mapsto y_1$ and point $x_2\mapsto y_2$ where $x_1 \in
P_1$ and $x_2\in P_2$; $P_1$, $P_2$ being P-sets. Let $y_1 \in
P_1'$ and $y_2\in P_2'$ with $P_1'$ and $P_2'$ being the images of
$P_1$ and $P_2$ under $\phi$. Then, the canonical distance ``d''
between $P_1$ and $P_2$ can evidently change under the action of
$\phi$ continuously (with respect to the corresponding Polish
topologies).

A Borel automorphism of $(\mathbb{B},\mathcal{B})$ then induces an
associated transformation of $(\mathbb{B}, d)$, say, to
$(\mathbb{B}, d')$, and that ``moves'' P-sets about in
$\mathbb{B}$, since (suitably defined) distance between the P-sets
can change under that Borel automorphism.

We call this the ``physical'' distance separating P-sets (as
extended bodies). We also (naturally) define distance separating
objects.

Now, the Borel automorphisms of $\mathbb{B}$ can be classified as
follows:
\begin{description} \item{(1)} those which {\em preserve\/} measures defined
on a specific P-set and \item{(2)} those which {\em do not
preserve\/} measures defined on a specific P-set \end{description}

Note that we are restricting our attention to only a specific
P-set/Object and not every P-set/Object is under consideration
here.

Measure-preserving Borel automorphisms then ``transform'' a P-set
maintaining its characteristic classes of (Lebesgue) measures on a
P-set, its physical properties.

Non-measure-preserving Borel automorphisms change the
characteristic classes of Lebesgue measures (physical properties)
of a P-set while ``transforming'' it. Evidently, such
considerations also apply to objects.

It is therefore permissible that a particular {\em periodic Borel
automorphism\/} leads to an {\em oscillatory motion\/} of a P-set
or an object while preserving its class of characteristic
measures.

We can then think of an object undergoing periodic motion as a
(physically realizable) time-measuring clock. Such an object
undergoing oscillatory motion then ``measures'' the time-parameter
of the corresponding (periodic) Borel automorphism since the
period of the motion of such an object is precisely the period of
the corresponding Borel automorphism.

Then, within the present formalism, a {\em measuring clock\/} is
therefore any P-set or an object undergoing {\em periodic motion}.

In the physical world, we do measure distances and construct
clocks in this manner. Then, crucially, the present formalism
represents measuring apparatuses, measuring rods and measuring
clocks, on par with every other thing that the formalism intends
to treat.

Such considerations then suggest an appropriate distance function,
{\em physical distance}, on the family of all P-sets/objects of
the space $(\mathbb{B},d)$. More than one such distance function
will be definable, depending obviously on the collection of P-sets
or objects that we may be considering in the form of a measuring
rod or measuring clock.

This above is permissible since we are dealing here with a
continuum which is a standard Borel space with Polish topology.
Relevant mathematical results can be found in \cite{trim6}.

A Borel automorphism of $(\mathbb{B},\mathcal{B})$ may change the
physical distance resulting into ``relative motion'' of objects.
We also note here that the sets invariant under the specific Borel
automorphism are characteristic of that automorphism. Hence, such
sets will then have their distance ``fixed'' under that Borel
automorphism and will be stationary relative to each other.

On a different note, an automorphism, keeping invariant a chain of
objects separating two other objects, can describe the situation
of two or more relatively stationary objects.

Effects of the Borel automorphisms of $(\mathbb{B},\mathcal{B})$
on the physical distance are then motions of physical bodies.
Furthermore, various physical phenomena will then be
manifestations of relevant properties of such automorphisms.

As an example, a joint manifestation of Borel automorphisms of the
space $(\mathbb{B}, \mathcal{B})$ and the association of a Dirac
$\delta$-distribution of an integrated measure with the points of
a P-set is a candidate reason behind Heisenberg's indeterminacy
relations in the present continuum formulation \footnote{Recall
that Einstein \cite{ein1} regarded the correctness of Heisenberg's
indeterminacy relations as being ``finally demonstrated''.}.
However, details regarding these considerations are outside the
limits of the present article.

But, intuitively, let it suffice to say that as the size of the
P-set gets smaller and smaller we ``know'' the position of the
point-particle (of integrated characteristics of a P-set) more and
more accurately. (But, recall that a P-set is never a singleton
subset of $\mathbb{B}$. So, complete positional localization is
not permissible.)

Now, that P-set ``transforms'' as a result of our efforts to
``determine any of its characteristic measures'' since these
``efforts or experimental arrangements'' are also Borel
automorphisms, not necessarily the members of the class of Borel
automorphisms keeping  invariant that P-set (as well as the class
of its characteristic measures).

Hence, any Borel automorphism (as an experimental arrangement)
purporting to ``determine'' a characteristic measure of that P-set
changes, in effect, the very quantity that it is trying to
determine. This peculiarity of the present continuum description
then leads to Heisenberg's (corresponding) indeterminacy relation.

Then, in the present continuum description, it is indeed possible
to explain the origin of Heisenberg's indeterminacy relations. The
present continuum description provides us therefore an origin of
indeterminacy relations, an alternative to their probabilistic
origin.

This circumstance is then extremely encouraging indeed.
Essentially, it tells us that one of the fundamental
characteristics, Heisenberg's indeterminacy relations, of the
theory of the quantum has an, indeed plausible, explanation in a
general relativistic theory of the continuum!

Notice then that, in the present considerations, we began with
none of the fundamental considerations of the concept of a
quantum. But, the present continuum formalism unfolds itself
before us in such a manner that one of the basic characteristics
of the conception of a quantum emerges out of the present
formalism.

Notice also that, in the present continuum description, we have
essentially done away with the ``singular nature'' of the
particles and, hence, also with the unsatisfactory dualism of the
field and the source particle. Furthermore, we have,
simultaneously, well-defined laws of motion (Borel automorphisms)
for the field and also for the well-defined conception of a
particle (of integrated measure characteristics).

Any particulate character or undulatory character perceptible in a
given physical phenomenon is then attributable to properties of
Borel automorphisms of the space $\mathbb{B}$, more precisely to
an interplay of Borel automorphisms simultaneously acting on the
space $\mathbb{B}$.

We refrain here from elaborating further on this issue. However,
intuitively, the particulate nature would be perceptible in a
phenomenon if the classical newtonian concepts ({\em eg.},
momentum) hold in some useful way. Otherwise, the undulatory
nature of the (total) field would be perceptible in that
phenomenon. Any phenomenon is, of course, a result of the
simultaneous interplay of Borel automorphisms of $\mathbb{B}$
here.

All these achievements of the present general relativistic
(continuum) formulation cannot be without any element of the
finality then.

The contention here is then the following: that the set of classes
of (Lebesgue) measures on P-sets of $(\mathbb{B},d)$ (as various
attributes of a physical particle) and the group of Borel
automorphisms of $(\mathbb{B},\mathcal{B})$ (resulting into
dynamics of P-sets) are, both, sufficiently large as to encompass
the entire diversity of physical phenomena.

In a definite sense, this approach then provides the theory of the
{\em total field\/} of Einstein's conception \cite{ein1}. It is
therefore also a continuum theory of everything in that sense.

The question naturally arises as to where, in the present
formulation, are the various constants of Nature such as Newton's
constant of gravitation, Planck's constant, constant speed of
light etc. Here, we only note that these constants arise from
relations of physical conceptions definable in this formulation.
In some definite sense, this present situation is describable as
obtaining {\em Constants without Constants}.

As an example, we need to analyze here as to how one obtains
Newton's law of gravitation in the present formalism in order to
obtain Newton's constant of gravitation. For this, we will need to
consider the ``gravitational mass'' measure of P-sets, and also
analyze the tendency of this P-set to oppose a change in its state
of motion, its inertia as per the conception of Galileo.

Any such analysis is, of course, based on the appropriate subgroup
of the group of Borel automorphisms of $(\mathbb{B}, \mathcal{B})$
as well as the association of a Dirac $\delta$-distribution (of an
``averaged measure'') with a P-set. Such considerations are, of
course, beyond the scope of the present article.

However, some definite conclusions are obtainable from very
general considerations of the present formalism and, we now turn
to a few such considerations.

Now, a Borel automorphism of $(\mathbb{B}, \mathcal{B})$ {\em
cannot\/} lead to a {\em singularity\/} of $\mathbb{B}$ and,
hence, to any kind of ``naked singularities'' from which some {\em
null\/} trajectory would reach other points of the manifold. Since
the present formalism is logically compelling, this would then
mean that the naked singularities would not be obtainable in the
Physics without a field-particle dualism.

Evidently, the Universe also does not originate or end in any
singularity as there also cannot be any such {\em singularity\/}
of the space $\mathbb{B}$ in this framework. But, an Expanding
Universe or a portion of $\mathbb{B}$ thereof, that may be {\em
hotter in the past than today\/} but having no {\em origin},
singular or otherwise, is thinkable in it.

This can be seen from the fact that certain Borel automorphisms
may move the P-sets about in the space $\mathbb{B}$ in such a
manner that these P-sets may move away from each other, while
simultaneously pushing some other P-sets closer to each other.
This is a very complicated picture but definitely thinkable
nonetheless.

Moreover, Borel automorphisms of $(\mathbb{B}, \mathcal{B})$ form
a group. Thus, we can always cross any 2-surface both ways. To see
this intuitively, let a Borel automorphism of $(\mathbb{B},
\mathcal{B})$ produce motion of an object ``into'' the given
2-surface. However, the inverse of that Borel automorphism,
producing motion pulling that object ``out'' of that 2-surface,
exists always is the point here.

Therefore, there does not arise any one-way membrane in the
present formalism. It therefore does not seem that the concept of
a black hole is any relevant to this Physics without the
field-particle dualism. As noted earlier, the concept of a black
hole is a child of the ``ghostly'' equations of the pure
gravitational field.

Now, a P-set, evidently, can be of any size and, hence, objects
can also be of any size. Therefore, a one-way membrane (black
hole) is not but, ultra-compact objects are of relevance to this
Physics without field-particle dualism.

As one more example, we consider Schr\"{o}dinger's cat paradox. In
the present formalism, some Borel automorphism of base space
$\mathbb{B}$ ``causes'' the disintegration of the atom while the
``measurement'' of this time-instant can take place using the
``suitable structure'' on the family of the P-sets of
$\mathbb{B}$. Therefore, the time-instant of disintegration of a
radioactive atom is ``well-defined as a parameter'' of that Borel
automorphism and, simultaneously, there is also the
Heisenberg-type indeterminacy involved in its measurement.

There is then some kind of Two-Time formalism in consideration
here. The time of disintegration of a radioactive atom as the
parameter of a Borel automorphism of $\mathbb{B}$ can then be {\em
different\/} than that which is {\em determinable}.

Recall that we began with no considerations of the quantum
conceptions. The disintegration of a radioactive atom is a quantum
consideration and the probabilistic interpretation leads us to a
``fuzzy'' description of macroscopic system - a cat. That a
possible resolution naturally arises here is then remarkable
indeed.

We now turn to some other aspects which were so beautifully
explained and elucidated by von Laue \cite{laue} - those related
to conservation principles in physical theories.

As von Laue expressed \cite{laue} it: {\em Mass is nothing but a
form of energy which can occasionally be changed into another
form. Up to now our entire conception of the nature of matter
depended on mass. Whatever has mass, - so we thought -, has
individuality; hypothetically at least we can follow its fate
throughout time. But, this does not hold for the elementary
particles.}

These remarks are then understandable in the present formalism
since the individuality of particles is that of the P-sets.

In the present formalism, a particle is a point of the P-set of
the space $\mathbb{B}$ with associated integrated measures defined
on that P-set. As a Borel automorphism of the space $\mathbb{B}$
changes that P-set, the integrated properties also change and,
hence, the initial particle changes into another particle(s),
since integrated measures change.

Of course, we then need to discover various laws of such
transformations of particles into one another in the present
formalism. But, it is clear at the outset that these will
crucially depend on the structure of the group of Borel
automorphisms of the space $\mathbb{B}$.

Von Laue concluded \cite{laue} his article with the following
relevant remarks: {\em ... Can the notions of momentum and energy
be transferred into every physics of the future? The uncertainty
relations of W. Heisenberg according to which we cannot precisely
determine location and momentum of a particle at the same time - a
law of nature precludes this -, can, for every physicist who
believes in the relation of cause and effect, only have the
meaning that at least one of the two notions, location and
momentum, is deficient for a description of the facts. Modern
physics, however, does not yet know any substitute for them.}

We may then conclude the present section with the following
remarks.

Here, the notion of energy is then that of the integrated measure
defined on a P-set of the space $\mathbb{B}$. The notion of the
momentum of a particle (as a point of the P-set of $\mathbb{B}$
with associated integrated measures on that P-set) is then that of
the appropriately defined notion of the rate of change of the
physical distance under the action of a Borel automorphism of
$\mathbb{B}$, including evidently any changes that may occur to
measures definable on that P-set. Therefore, the notions of energy
and momentum of a particle are certainly (well-) definable in the
present formalism.

Further, none of these two notions is any deficient for a
description of the facts since Heisenberg's indeterminacy
relations are also ``explainable'' within the present formalism.
This explanation crucially hinges on the fact that the points of
the space $\mathbb{B}$ can never be particles since these, as
singleton subsets of the space $\mathbb{B}$, are never the P-sets.
It is only in the sense of associating the measures integrated
over a P-set that the points of the space $\mathbb{B}$ are
particles.

The measurable location of a particle is essentially a {\em
different\/} conception and that depends on the physical distance
definable on the class of all P-sets of the space $\mathbb{B}$.
The measurable momentum of a particle is also dependent on the
notion of the physical distance changing under the action of a
Borel automorphism of $\mathbb{B}$.

Therefore, it does seem possible to combine the virtues of the
theory of the quantum and of the general theory of relativity in a
formalism that does possess the notions of energy and momentum,
both. Then, this formalism will further unite more number of
conservation laws than those already united in special relativity.

Now, as seen earlier, physical processes occur as a result of the
(combined effects of) the Borel automorphisms of the space
$\mathbb{B}$ and these include the processes of quantum character.
Then, it is thinkable that ``some suitable statistical
description'' is obtainable by ``approximating'' the effects of
these Borel automorphisms acting on $\mathbb{B}$ to produce the
concerned physical processes of quantum character. This,
intuitively speaking, can be considered as the probabilistic
description of the involved physical processes.

Hence, at this point, we also note that, in relation to the
present formalism, the theory of the quantum as represented by
Schr\"{o}dinger's $\Psi$-function can be expected to assume a
place which is similar to that of the usual statistical mechanics
within the realm of the classical newtonian theory. (Details of
these considerations are, once again, outside the scope of the
present article.)

Provided that the description of physical systems of Nature based
on a continuum is permissible, the approach \cite{smw01} followed
here in Section \ref{totalfield} is also a logically compelling
approach as this article discussed.

Of course, many mathematical, physical details and their
implications need to be worked out before we can ``test'' the
above described theoretical framework vis-a-vis experimentation or
astronomical observations. But, confidence may be voiced that it
will stand tests of experimentations and/or observations since it
is a logically compelling approach as discussed here.

A comment on the mathematical methods would not be out of place
here. Then, we note that the mathematical formalism of the ergodic
theory is what is of immediate use for the physical framework of
the present section. This much is already clear from the present
considerations.

However, it is not entirely satisfactory to use the present
methods of ergodic theory. One of the primary reasons for this
state of affairs is the inability of the present methods in
ergodic theory to let us handle, in a physical sense, the P-sets.
Some newer methods are then required here.

\section{Concluding remarks}
In the present article, we outlined the reasons behind the
logically compelling character of the approach adopted in Section
\ref{totalfield}.

In doing so, we also critically examined the reasons behind the
``failures'' of other approaches to General Relativity based on
the equations of the pure gravitational field and Einstein's
makeshift equations for the matter fields.

In particular, the equations of the pure gravitational field are
``ghostly'' in the sense of Newton's Absolute Space or in the
sense of this approach providing only the equations for the field
without any possibility for the equations of motion for the
sources of the field. These equations cannot therefore provide the
means to verify or test predictions in any manner whatsoever.

Then, any solution of these field equations of the pure
gravitational field is ``ghostly'' as well. Any conclusion of a
physical nature obtained using these equations is therefore
dubious.

Further, Einstein's makeshift equations for general field are
based on an ill-defined concept of the energy-momentum tensor.
This is mainly because various concepts leading us to the notion
of the energy-momentum tensor are based primarily on the concept
of a point-particle which is not defined {\em abinitio\/} in a
(dynamical) geometric approach such as the one postulated and
adopted by general relativity. Clearly, a point-particle as a
spacetime singularity leads to the impossibility of the definition
of the energy-momentum tensor.

The pivotal problem with the solutions of the makeshift field
equations is then that of their physical interpretation in view of
some continuum description of physical systems. We cannot consider
these solutions as describing some ``fluid'' matter since the
conceptions of fluid properties are ill-defined to begin with.
Therefore, it follows that not all the solutions of Einstein's
makeshift field equations are necessarily useful for the continuum
description of physical bodies. Of particular mention is the
Friedmann-Lemaitre-Robertson-Walker (FLRW) geometry of the
standard model of the big bang cosmology.

(Therefore, we must, unassumingly, also seek the pardon of all
those whose sincere and herculean efforts provided us the many
solutions \cite{stdworks} of Einstein's makeshift field equations.
Clearly, these equations are based on ill-defined considerations.
Furthermore, the solutions of these highly non-linear partial
differential equations do not follow any superposition principle.
Therefore, it also follows that even if there existed some
particular spacetime geometry useful for some specific continuum
description of physical systems, it would not be a superposition
of other spacetime geometries. Other spacetime geometries are then
rendered unphysical in this sense.)

However, it also follows that only the smooth spacetime geometries
are the ones that need to be considered if the description of
physical systems is permissible on the basis of a continuum. The
question then arises of choosing some particular spacetime
geometry that can provide us this continuum description of
physical systems.

Clearly, the answer to this question requires use of some physical
principles, not contained within the framework of standard general
relativity (the framework of Einstein's makeshift equations) since
this theory permits many different smooth spacetime geometries
containing matter fields. This is what we considered in Section
\ref{totalfield}.

In Section \ref{quantumtheory}, somewhat separated section from
the earlier considerations, we then considered developments
related to the theory of the quantum conception. In this section,
we recalled Einstein's exposition of the relevant developments of
the concept of a quantum. In particular, we noted the precise
nature of problems with the (classical) newtonian theories. We
also noted that the contradiction with the laws of newtonian
mechanics is of more fundamental nature than that with the laws of
Maxwell's electromagnetism.

The theory of the quantum is based primarily on the probabilistic
conceptions. Schr\"{o}dinger's $\Psi$-function then provides us
only the probability of any physical event. As is well known, this
then leads to an indeterminacy in the simultaneous measurement of
canonically conjugate (classical) variables. This is the
probabilistic origin of Heisenberg's indeterminacy relations.

This probabilistic nature of the theory of the quantum leads to
paradoxical situations of serious concern if every physical
system, microscopic or macroscopic notwithstanding, obeyed
probabilistic laws. We then also considered, in Section
\ref{criticismofquantum}, various objections of such serious
nature related to this interpretation of the theory of the
quantum. In particular, some of the serious objections raised by
Einstein \cite{ein1} were quoted in details.

In view of the considerations of these objections, it is then
possible to adopt the view that the theory of the quantum,
Schr\"{o}dinger's $\Psi$-function, represents an ensemble(s) of
systems.

With this above point of view, the theory of the quantum may then
be expected to assume, in relation to an appropriate general
relativistic theory, one based on the principle of general
covariance, for the case (b) of general fields, a place similar to
that of the statistical mechanics within the realm of the
classical newtonian framework. Then, various paradoxical
situations arising in the theory of the quantum as embodied in
Schr\"{o}dinger's $\Psi$-function evidently disappear. Einstein
\cite{ein1} had, very clearly, perceived this situation with the
theory of the quantum.

With this view point, within the theory of the quantum as embodied
in Schr\"{o}dinger's $\Psi$-function, there is obviously no
possibility of obtaining any non-probabilistic theoretical
framework for the description of physical phenomena. Einstein had
also clearly recognized this aspect.

Therefore, we had to look ``elsewhere'' for the theoretical
framework (obeying the principle of general covariance) that would
provide us the theory of the case (b) of general fields. This is
what Einstein meant in his remarks which were quoted in the
context of the relevant discussions.

Therefore, on the basis of some few (physically reasonable)
principles (observed to hold in Nature), we developed the approach
\cite{smw01} outlined in Section \ref{totalfield}. This
theoretical framework then obeys the principle of general
covariance, in a round about manner, since the group of Borel
automorphisms of the space $\mathbb{B}$ is {\em very large}.

We then saw that this formalism provides us a clear possibility of
explaining the laws of the quantum realm while simultaneously
treating the concept of a particle in a non-singular manner. It
also shows us the possibility of visualizing the theory of the
quantum as embodied in Schr\"{o}dinger's $\Psi$-function to be of
similar character to the usual statistical mechanics.

In particular, the (continuum) formalism of Section
\ref{totalfield} provides \cite{smw03} a non-probabilistic
explanation for Heisenberg's indeterminacy relations. Furthermore,
it also allows us satisfactory resolutions of different (serious)
paradoxical situations faced by the theory of the quantum as
embodied in Schr\"{o}dinger's $\Psi$-function.

Moreover, it is then also clear that many of the current ideas in
theoretical cosmology need drastic modifications. Notably, there
is no singularity of the space(time) in the past as well as in the
future in the present framework.

The physical matter in the universe can only be ``rearranged'' in
the present framework (of Section \ref{totalfield}) that is a
complete field theory. This fact can be expected to have important
implications and consequences \cite{smw06} for our understanding
of the cosmological phenomena.

So also our models of some (galactic as well as extragalactic)
high energy sources need drastic changes. There does not arise a
black hole or a naked singularity in the present framework.
Therefore, our models of astronomical sources cannot be based on
these conceptions.

In summary, Einstein had been one of the proponents of the
ensemble interpretation of the quantum theory. We then recall here
that, to emphasize Einstein's contributions to developments in the
theory of the quantum, Max Born wrote \cite{ein1} about Einstein
that: {\em He has seen more clearly than anyone before him the
statistical background of the laws of physics, and he was a
pioneer in the struggle for conquering the wilderness of quantum
phenomena. Yet later, when out of his own work, a synthesis of
statistical and quantum principles emerged which seemed to be
acceptable to almost all physicists, he kept himself aloof and
sceptical. Many of us regard this as a tragedy - for him, as he
gropes his way in loneliness, and for us who miss our leader and
standard-bearer.}

However, as a result of our combined efforts,  with so many in the
past wrestling with and weakening to a large extent the difficult
problems of grasping the Nature, we have been able to see a little
beyond their vision.

It was, of course, extremely difficult for the physicists of
earlier times, Einstein and others included, to imagine the
current path \cite{smw01}, the formalism of the Section
\ref{totalfield}, in those times, turbulent times of vigorous
theoretical and experimental activities. In fact, many of the
mathematical conceptions of Section \ref{totalfield}, specifically
those of the ergodic theory, were not even available to them. But,
from our present considerations, it then should be clear that
Born's leader, Einstein, had indeed the right intuition all along.

The physically complete framework of the Section \ref{totalfield}
is also in conformity with the relevant ideas of Descartes
\cite{ein-pop, smw03}. Clearly, this field-theoretic program of
Section \ref{totalfield} is then also: the complete description of
any (individual) real situation (as it supposedly exists
irrespective of any act of observation or substantiation).

Then, a stage can be said to have been certainly reached in the
history of Physics, once again since Newton's times, in that we
have a physically complete framework in the formalism of Section
\ref{totalfield} for describing the totality of physical phenomena
in Einstein's sense.

\end{document}